\documentclass[11pt]{article}
\usepackage{graphicx,amssymb}
\usepackage{citesort}

\input{text/cteq6.fig}
\input{text/cteq6.pre}

\begin{document}


\begin{titlepage}

\noindent
\begin{tabular}{l}
\DATE
\end{tabular}
\hfill
\begin{tabular}{l}
\PPrtNo
\end{tabular}

\vspace{1cm}

\begin{center}
\renewcommand{\thefootnote}{\fnsymbol{footnote}}
{
\LARGE \TITLE
}

\vspace{1.25cm}
{\large  \AUTHORS}

\vspace{1.25cm}

\INST
\end{center}

\vfill

\ABSTRACT                 

\vfill

\newpage
\end{titlepage}

\setcounter{footnote}{0}
\tableofcontents

\section{Introduction}

\label{sec:Intro}

Progress on the determination of the parton distribution functions (PDF's) of
the nucleon, from global quantum chromodynamics (QCD) analysis of hard
scattering processes, is central to precision standard model (SM)
phenomenology, as well as to new physics searches, at lepton-hadron and
hadron-hadron colliders. There have been many new developments in recent
years, beyond the conventional analyses that underlie the widely used PDF's
\cite{Mrst0,Cteq,GRV}. These developments have been driven by the need to
quantify the uncertainties of the PDF's and their physical predictions
\cite{Alekhin,Botje,ZomerDis,GKK,Bialek,Ball,Multivariate,Hesse,Lagrange}. We
report in this paper on a comprehensive new global QCD analysis based on the
most current data, and on the recently developed methods of uncertainty study
of \cite{Multivariate,Hesse,Lagrange}. This new analysis includes a full
treatment of all available correlated experimental errors, as well as an
extensive exploration of the parametrization of the input nonperturbative
PDF's.

Although this work is built on the series of previous CTEQ parton
distributions \cite{Cteq}, it represents more than an evolutionary updating
of previous work to incorporate new experimental data sets. The methodology
of \cite{Multivariate,Hesse,Lagrange} goes beyond the traditional paradigm of
producing some subjectively chosen ``best fits.'' It introduces a set of
efficient and practical tools to characterize the parton parameter space in
the neighborhood of the global minimum. This allows the systematic
exploration of the uncertainties of parton distributions and their physical
predictions due to known experimental errors and due to the input theoretical
model parameters.

There are many complex issues involved in a comprehensive global parton
distribution analysis. Foremost among these on the experimental side is the
``imperfection'' of real experimental data compared to textbook
behavior---for instance, some experimental measurements appear to be
statistically improbable because the $\chi^2\:/\:N$ deviates from $1$
substantially more than the expected $\pm\sqrt{2/N}$; or different precision
experimental measurements of the same physical quantities appear to be
statistically incompatible in all regions of the model parameter space. The
methods of \cite{Multivariate,Hesse,Lagrange} cannot resolve these
problems---no global analysis method can---but the tools developed in this
formalism make it possible to look deeper into some of these problems, in
order to assess the acceptability and compatibility of the affected data sets
in more practical terms, and to suggest pragmatic ways to deal with the
apparent difficulties. These detailed studies were not possible in the
conventional analyses \cite{Mrst0,Cteq,GRV} and \cite{Mrst01}. On the
theoretical side, the uncertainties on the perturbative QCD (PQCD)
calculations of the various physical processes included in the global
analysis are not easily quantified in a uniform way. We do not address that
problem here.

Section \ref{sec:NewInput} summarizes the experimental and theoretical input
to the global analysis, emphasizing the new elements in data sets and in
methodology. Section \ref{sec:PdfRes} presents results on the new generation
of PDF's: the ``standard'' CTEQ6 sets, as well as the eigenvector sets that
characterize the uncertainties. It also includes discussions of various
physics issues relating to these global fits. Section \ref{sec:PhyPred}
presents some general results on uncertainties of physics predictions due to
PDF's, in the form of parton-parton luminosity functions for the Tevatron
RunII and for the LHC. Section \ref{sec:Compare} compares this work to
previous studies of PDF uncertainties \cite{Alekhin,Botje,ZomerDis,GKK}
(primarily on DIS experiments), and to a recent general PDF analysis \cite
{Mrst01}. Four appendices concern: (\ref{app:C6m}) detailed information on
the new standard parton distribution set CTEQ6M; (\ref{app:ErrorAna}) new
formulas and tools for understanding the significance of $\chi^2$ values for
individual experiments and estimating ranges of uncertainties in the
comparison of fits with data; (\ref{app:HT}) assessing the need for
higher-twist (power-law) contributions to the theory model; and
(\ref{app:Param}) studying the impact of the flexibility associated with the
choice of parametrization for the nonperturbative PDF's.

\section{New Input and Methodology}

\label{sec:NewInput}

The next-to-leading order (NLO) global QCD analysis carried out in this work
is built on the same basis as the previous CTEQ parton distribution sets
\cite{Cteq}. In this section, we describe the new experimental input and the
new theoretical techniques, which together have enabled substantial progress
in this generation of global analysis.

\subsection{Experimental Data Sets}

\label{sec:NewData}

Since the CTEQ5 analysis, many new experimental data sets have become
available for an improved determination of parton distributions. Particularly
noteworthy are the recent neutral current deep-inelastic scattering (DIS)
structure function measurements of H1 \cite{H1} and ZEUS \cite{ZEUS}, and the
inclusive jet cross section measurement of D\O\ \cite{D0jet} (in several
rapidity bins, up to a rapidity of 3). The greater precision and expanded
$(x,Q)$ ranges compared to previous data in both processes provide improved
constraints on the parton distributions. Other recent data used in the
analysis are the updated E866 measurements of the Drell-Yan deuteron/proton
ratio \cite{E866} and the re-analyzed CCFR measurement of $F_2$ \cite{CCFR2}.

These new results complement the fixed-target DIS experiments of BCDMS
\cite{BCDMSp,BCDMSd}, NMC \cite{NMC}, CCFR $F_3$ \cite{CCFR3}, the Drell-Yan
measurement of E605 \cite{E605}, the CDF measurement of $W$-lepton asymmetry
\cite{Wasymm}, and the CDF measurement of inclusive jets \cite{CDFjet}, which
this study shares with the earlier CTEQ5 analysis. Even for these older
experiments, our new analysis, by including correlated systematic errors
(cf.\ Sec.~\ref{sec:SysErr} and Appendix~\ref{app:SysErr}), incorporates more
details than previous \cite{Mrst0,Cteq} and recent \cite{Mrst01} analyses.
For instance, for the BCDMS and NMC experiments we now use the data sets
measured at separate energies (which contain full information on correlated
errors) instead of the combined data sets (which, due to re-binning of data
points, retain only some effective point-to-point uncorrelated errors of
uncertain statistical significance).

\subsection{$\protect{\mathbf{\chi^2}}$
Function and Treatment of Correlated Systematic Errors}

\label{sec:SysErr}

The very extensive and precise DIS data from fixed-target and HERA
experiments provide the backbone of parton distribution analysis. In order to
make full use of the experimental constraints, it is important to incorporate
the available information on correlated systematic errors
\cite{Alekhin,Botje,ZomerDis,GKK,Multivariate,Hesse,Lagrange}. The same is
true for recent data on inclusive jets \cite{D0jet,CDFjet}, where the
experimental uncertainties are dominated by systematics. Thus, for the first
time in the CTEQ series of analyses, we have included the correlated errors
wherever available. Our analysis is simplified by a novel way of treating
systematic errors, formulated in \cite{Lagrange} (cf.~also the recent review
\cite{BotjeRev}). As an introduction to subsequent discussions based on this
method, we mention its basic ideas and main features here. A more complete
summary of this method is given in Appendix \ref{app:SysErr}, as needed to
explain the detailed comparison between data and fits discussed in that
section.

In global fits not including correlated errors, one would minimize a naive
global $\chi ^{2}$ function defined simply as $\chi
_{0}^{2}=\sum_{\mathrm{expt.} }\sum_{i=1}^{N_{e}}\left( D_{i}-T_{i}\right)
^{2}/\sigma _{i}^{\prime {2}}$, where $N_{e}$ is the number of data points in
experiment $e$, $D_{i}$ is a data value, $T_{i}$ is the corresponding theory
value (which depends on the PDF model), and $\sigma _{i}^{\prime 2}=\sigma
_{i}^{2}+\Sigma_{i}^{2}$ is
the statistical ($\sigma _{i}^{2}$) and point-to-point systematic ($%
\Sigma_{i}^{2}$) errors added in quadrature. The function $\chi_{0}^{2}$
provides the simplest means to search for candidates of ``good'' global
fits, but has rather limited use in assessing uncertainties of the resulting
fits.

If there are $K$ sources of correlated systematic errors, specified by
standard deviations $\{\beta _{1i},\beta _{2i}, \dots ,\beta _{Ki}\}$, in
addition to an uncorrelated systematic error $u_{i}$ for data point $i$, then
a standard method to improve the treatment of experimental errors is to
construct the covariance matrix %
$V_{ij}=\alpha_{i}^{2}\delta _{ij} +\sum_{k=1}^{K}\beta _{ki}\beta _{kj}$, %
from which a global $\chi ^{2}$ function is defined by %
$\chi^{2}=\sum_{\mathrm{expt.}}\sum_{i,j=1}^{N_{e}} \left( D_{i}-T_{i}\right)
V_{ij}^{-1}\left( D_{j}-T_{j}\right)$. %
(Here $\alpha_{i}^{2}=\sigma_{i}^{2}+u_{i}^{2}$.) %
An alternative method is to add $K$ parameters, each associated with one
systematic error, and to minimize an \emph{extended} $\chi^{2}$ function with
respect to the combined set of experimental and theoretical parameters
(denoted by $\chi'^2$ in Appendix \ref{app:SysErr}, Eq.~(\ref{eq:Chi2sys})).
These well-known methods are equivalent. Both face some practical, even
formidable, problems in the context of global QCD analysis because of the
large number of data points (which can make the inversion of the covariance
matrix numerically unstable) and the large number of fitting parameters
(which becomes unmanageable when all systematic errors from all experiments
are included).

The method formulated in \cite{Lagrange} overcomes these difficulties by
solving the problem of optimization with respect to the correlated systematic
errors analytically. The result is an effective $\chi ^{2}$ minimization
problem with respect to only the theory parameters, as for the simple case.
The resulting $\chi ^{2}$ function has the form (cf.\ %
Eq.~(\ref{eq:chi2corr})):
\begin{equation}
\chi ^{2}
=\sum_{\mathrm{expt.}}\ \left\{ \sum_{i=1}^{N_{e}}
\frac{\left(D_{i}-T_{i}\right)^{2}}{\alpha _{i}^{2}}
-\sum_{k,k^{\prime}=1}^{K_{e}}
B_{k}\left( A^{-1}\right)_{kk^{\prime}}B_{k^{\prime}}\right\},
\label{eq:SysChi1}
\end{equation}
where $\{B_{k}\}$ is a $K_{e}$-component vector, $\{A_{kk^{\prime }}\}$ is a
$K_{e}\times {K}_{e}$ matrix, and
\begin{equation}
B_{k}=\sum_{i=1}^{N_{e}}\frac{\beta _{ki}\left( D_{i}-T_{i}\right) }{\alpha
_{i}^{2}},\qquad A_{kk^{\prime }}=\delta _{kk^{\prime }}+\sum_{i=1}^{N_{e}}%
\frac{\beta _{ki}\beta _{k^{\prime }i}}{\alpha _{i}^{2}}.  \label{eq:SysChi2}
\end{equation}
Not only is the matrix inversion simpler and more stable ($K_{e}\times {K}%
_{e}$ with $K_e \leq 10$, compared to $N_{e}\times {N}_{e}$ with $N_e$ as
large as 300, in the correlation matrix approach), the explicit formula for
the contribution of the various sources of systematic errors in
Eq.~(\ref{eq:SysChi1}) provides a useful tool to evaluate the uncertainties
in detailed analysis of the fits---as explained in Appendix \ref{app:SysErr}
and used in discussions in subsequent sections---which is not available in
the traditional approach.

\subsection{Methods for Analysis of Uncertainties}

\label{sec:NewMethod}

The widely used parton distributions \cite{Mrst0,Cteq,GRV,Mrst01} are
obtained from global analyses using a ``best-fit'' paradigm, which selects
the \emph{global minimum} of the chosen $\chi^2$ function. However, the
``best'' PDF's are subject to various uncertainties. Within this paradigm,
the problem of estimating uncertainties of the PDF's can only be addressed in
an {\it ad hoc} manner, usually by examining alternative fits obtained by
subjective tuning of selected degrees of freedom. Recent efforts to assess
the uncertainties objectively, using established statistical methods, have
been mostly concerned with the precision DIS experimental data
\cite{Alekhin,Botje,ZomerDis,GKK}, rather than the global analysis of all
hard scattering data. As mentioned in the introduction, there are formidable
complications when standard statistical methods are applied to global QCD
analysis. The basic problem is that a large body of data from many diverse
experiments,%
\footnote{%
For our analysis, there are $\sim 1800$ data points from $\sim 15$ different
sets of measurements with very different systematics and a wide range of
precision.} %
which are not necessarily compatible in a strict statistical sense, is being
compared to a theoretical model with many parameters, which has its own
inherent theoretical uncertainties.

In recent papers \cite{Multivariate,Hesse,Lagrange}, we have formulated two
methods, the Hessian and the Lagrange, which overcome a number of
long-standing technical problems encountered in applying standard error
analysis to the complex global analysis problem. We are now able to
characterize the behavior of the $\chi^2$ function in the \emph{neighborhood}
of the global minimum in a reliable way. This provides a systematic method to
assess the compatibility of the data sets in the framework of the theoretical
model \cite{ColPump}, and to estimate the uncertainties of the PDF's and
their physical predictions within a certain practical tolerance. The basic
ideas are summarized in the accompanying illustration, adapted from
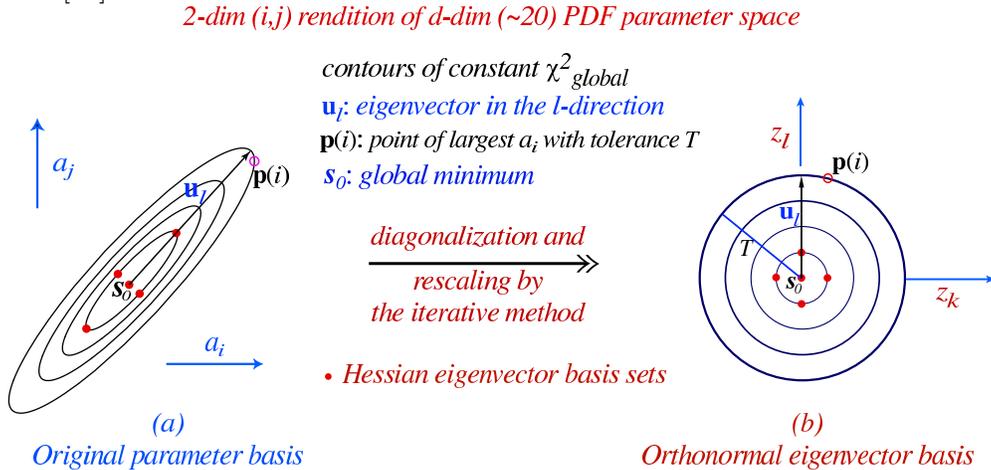
\cite{Hesse}:\vspace{0ex}
\newline%
\GraphHesse%
\newline %
The behavior of the global $\chi^2$ function in the
neighborhood of the minimum in the PDF parameter space is encapsulated in
$2N_p$ sets of eigenvector PDF's (where $N_p\sim 20$ is the number of free
PDF parameters), represented by the solid dots in the illustration.
These eigenvectors are obtained by an iterative procedure to
diagonalize the Hessian matrix, adjusting the step sizes of the numerical
calculation to match the natural physical scales. This procedure efficiently
overcomes a number of long-standing obstacles%
\footnote{%
The obstacles are due to difficulties in calculating physically meaningful
error matrices by finite differences, in the face of (i) vastly different
scales of eigenvalues ($\sim 10^{7}$) in different, a priori unknown,
directions in the high-dimension parameter space, and (ii) numerical
fluctuations due to (multi-dimensional) integration errors in the theoretical
(PQCD) calculation
and round-off errors.} %
encountered when applying standard tools to perform error propagation in the
global $\chi^2$ minimization approach. Details are given in
\cite{Multivariate,Hesse}.

The uncertainty analysis for our new generation of PDF's makes full use of
this method. The result is $2N_p+1$ PDF sets, consisting of the best fit
$S_0$ and eigenvector basis sets in the plus and minus directions along each
eigenvector. From these PDF sets we can calculate the best estimate, and the
range of uncertainty, for the PDF's themselves and for any physical quantity
that depends on them. The uncertainty can be computed from the simple master
formula
\begin{equation}
\Delta X= \frac{1}{2} \left( \sum_{i=1}^{N_{p}}
\left[ X(S_{i}^{+})-X(S_{i}^{-}) \right]^{2}
\right)^{1/2},
\label{eq:DeltaX}
\end{equation}
where $X$ is the observable, and $X(S_{i}^{\pm})$ are the predictions for $X$
based on the PDF sets $S_{i}^{\pm}$ from the eigenvector basis.

\subsection{Perturbative QCD Parameters and Input}

\label{sec:QcdParam}

The fundamental parameters of perturbative QCD are the coupling $\alpha_{s}$
and the quark masses. In principle, these parameters can be determined in the
global fit, along with the nonperturbative parton distribution functions. In
practice, they are determined more precisely and definitively in dedicated
measurements. We therefore treat these parameters as input to our global
analysis. For \mbox{$\alpha_{s}$}, we use $\alpha_{s}(M_Z)=0.118$, the
Particle Data Group average \cite{PDG}. (However, cf.\ the discussion on
leaving $\alpha_{s}$ free in the fitting in Sec.~\ref{sec:alpha} and
Sec.~\ref{sec:Mrst01}.) The light quarks ${u,d,s}$ are treated as massless,
as usual. In the \mbox{\small {$\overline {MS}$}}\ scheme, the evolution
kernels of the PDF's are mass-independent: charm and bottom masses enter only
through the scales at which the heavy quark flavors are turned on in the
evolution. We take $m_c=1.3$ GeV and $m_b=4.5$ GeV.

The hard matrix elements that enter into the PQCD calculations are all taken
to be at NLO (except for the LO fit, of course). Although it is now possible
to incorporate charm and bottom mass effects in the NLO hard cross sections,
in the so-called variable flavor number scheme (VFNS)
\cite{ACOT,TR,CSV,CFGN}, we have decided to use the conventional (zero-mass
parton) hard cross sections in our standard analysis for three practical
reasons. First, hard cross-section calculations in the VFNS (with non-zero
parton masses) are not yet available for processes other than DIS. Second,
even in DIS, there are different ways to implement the VFNS. This introduces
yet another type of ``scheme dependence'', and hence a source of confusion
for users of PDF's. (A recently proposed ``natural'' implementation of the
VFNS \cite{KST01}, if widely accepted, could alleviate this source of
confusion.) Finally, since most users of PDF's only have at their disposal
the standard hard cross sections for zero-mass partons, it is preferable to
have PDF's determined by a global analysis calculated in the same way as they
are used. In a subsequent paper, we will study the heavy quark mass effects
in global analysis in detail, and publish PDF's determined in the VFNS with a
full treatment of quark mass effects, using the natural implementation of
\cite{KST01}.

\subsection{Parametrization of Nonperturbative Input PDF's}

\label{sec:Parametrization}

The nonperturbative input to the global analysis are PDF's specified
in a parametrized form at a fixed low-energy scale $Q_{0}= 1.3\,
\mathrm{GeV}$.%
\footnote{%
In CTEQ5 and other recent global fits, the slightly lower value $Q_{0} =
1\,\mathrm{GeV}$ was used. Because QCD evolution is quite rapid in the $Q\sim
1\mathrm{-}2$\,GeV range, one gets enhanced sensitivity to the parameters
when starting from a relatively low scale. However, that rapid
variation, which occurs below the region of applicability of PQCD, can
produce unusual behavior of the functions at small $Q$, which may have no
\emph{physical} significance. Cf.\ the discussions in Sec.\
\ref{sec:Mrst01} and Appendix
\ref{app:Param}.} %
The particular functional forms, and the value of $Q_0$, are not crucial,
as long as the
parametrization is flexible enough to describe the range of behavior
permitted by the available data at the level of accuracy of the data.
Intuitive notions of smoothness may be built into the input functions. The
PDF's at all higher $Q$ are determined from the input functions by the NLO
perturbative QCD evolution equations.

The functional form that we use is
\begin{equation}
x \, f(x,Q_{0}) = A_{0} \, x^{A_{1}} \, (1-x)^{A_{2}} \, e^{A_{3} x} \, (1 +
e^{A_{4}}\:x)^{A_{5}}  \label{eq:param}
\end{equation}
with independent parameters for parton flavor combinations $u_{v} \equiv u -
\bar{u}$, $d_{v} \equiv d - \bar{d}$, $g$, and $\bar{u}+\bar{d}\,$. We assume
$s = \bar{s} = 0.2 \, (\bar{u} + \bar{d})$ at $Q_0$. The form (\ref
{eq:param}) is ``derived'' by including a 1:1 Pad\'e expansion in the
quantity $d\left[\log(xf)\right]/dx$. This logarithmic derivative has an
especially simple form for the time-honored canonical parametrization
$x\,f(x) = A_{0} \, x^{A_{1}} \, (1-x)^{A_{2}}$. For our parametrization
there are poles at $x = 0$ and $x=1$ to represent the singularities
associated with Regge behavior at small $x$ and quark counting rules at large
$x$, along with a ratio of (linear) polynomials to describe the intermediate
region in a smooth way.

Equation (\ref{eq:param}) provides somewhat more versatility than previous
CTEQ parametrizations. For some flavors, it has more freedom than needed, so
that not all of the parameters can be determined by the present data. In
those cases, some parameters are held fixed, guided by the Hessian method in
the final fitting process as discussed in Sec.\ \ref{sec:PdfUnc} and in
Ref.\ \cite{Hesse}. In the future, more data may find even this
parametrization too restrictive---first for the very well constrained
$u(x)$---and it will be natural to proceed to a 2:2 Pad\'e form. To
distinguish the $\bar{d}$ and $\bar{u}$ distributions, we parametrize the
\emph{ratio} $\bar{d}/\bar{u}$, as a sum of two terms:
\begin{equation}
\bar{d}(x,Q_{0})/\bar{u}(x,Q_{0}) = A_{0}\,x^{A_{1}}\,(1-x)^{A_{2}} \; +
\;(1+A_{3}\:x)\,(1-x)^{A_{4}} \;.  \label{eq:dou}
\end{equation}
Altogether we use 20 free shape parameters to model the PDF's at $Q_0$.

An extensive study of the sensitivity of the global fits
to the choice of parametrization has been carried out in the
process of this analysis, by trying different parametrizations, and by using
different values of $Q_0$
(which is equivalent to changing the functional forms).
In the main body of this paper, we concentrate on results
obtained with the standard choices described above.
Comments on the effects of parametrization on the physics results
will be made in the text as appropriate.
Some studies of results obtained with alternative parametrizations
are described in Appendix~\ref{app:Param}.

\section{Results on New Parton Distributions}

\label{sec:PdfRes}

With the theoretical and experimental input, methods, and parametrizations
described above, we arrive at a \emph{standard set} of parton distributions
(the nominal ``best fit'') together with a \emph{complete set of eigenvector
parton distribution sets} that characterizes the neighborhood of acceptable
global fits in the parton parameter space.
The study is carried out mainly in the $\overline{MS}$ scheme.\footnote{%
For the convenience of certain applications, we also present one standard
set each of parton distributions in the DIS scheme and at leading order.
Cf.\ Sec.~\ref{sec:DisLo}.} We now discuss the salient features of the
results and the related physics issues.

\subsection{The New Standard PDF Sets}

\label{sec:BestFit}


The standard set of parton distributions in the \mbox{\small {$\overline
{MS}$}} scheme, referred to as CTEQ6M, provides an excellent global fit to
the data sets listed in Sec.~\ref{sec:NewData}. An overall view of these
PDF's is shown in Fig.~\ref{fig:PDFview}, at two scales $Q=2$ and $100$ GeV.
The overall $\chi ^{2}$ for the CTEQ6M fit is $1954$ for $1811$ data points.
The parameters for this fit and the individual $\chi^{2}$ values for the data
sets are given in Appendix \ref{app:C6m}. In the next two subsections, we
discuss the comparison of this fit to the data sets, and then describe the
new features of the parton distributions themselves. Quantitative comparison
of data and fit is studied in more depth in Appendix \ref{app:ErrorAna}%
\figPDFview %

\subsubsection{Comparison with Data}

\label{sec:CompData}

The fact that correlated systematic errors are now fully included in the
fitting procedure allows a more detailed study of the quality of fits than
was possible in the past. We can take the correlated systematic errors into
account explicitly when comparing data and theory, by using the procedure
discussed in Sec.~\ref{sec:Tools} of Appendix~\ref{app:ErrorAna}. In
particular, based on the formula for the extended $\chi^2$ function expressed
in the simple form Eq.~(\ref{eq:Chi2sys1}), we obtain a precise graphical
representation of the quality of the fit by superimposing the theory curves
on the shifted data points \{$\widehat{D}_i$\} containing the fitted
systematic errors. The remaining errors are purely uncorrelated, hence are
properly represented by error bars. We use this method to present the results
of our fits whenever possible.


Figure \ref{fig:Hone} shows the comparison of the CTEQ6M fit to the latest
data of the H1 experiment \cite{H1}. The extensive data set is divided into
two plots: (a) for $x < 0.01$, and (b) for $x > 0.01$. In order to keep the
various $x$ bins separated, the values of $F_{2}$ on the plot have been
offset vertically for the $k$th bin according to the formula: $
\mathrm{ordinate}\, = F_{2}(x,Q^{2}) + 0.15\:k$. The excellent fit seen in
the figure is supported by a $\chi^2$ value of $228$ for $230$ data points.
Similarly, Fig.~\ref{fig:ZEUS} shows the comparison to the latest data from
ZEUS \cite{ZEUS}. One again sees very good overall agreement. %
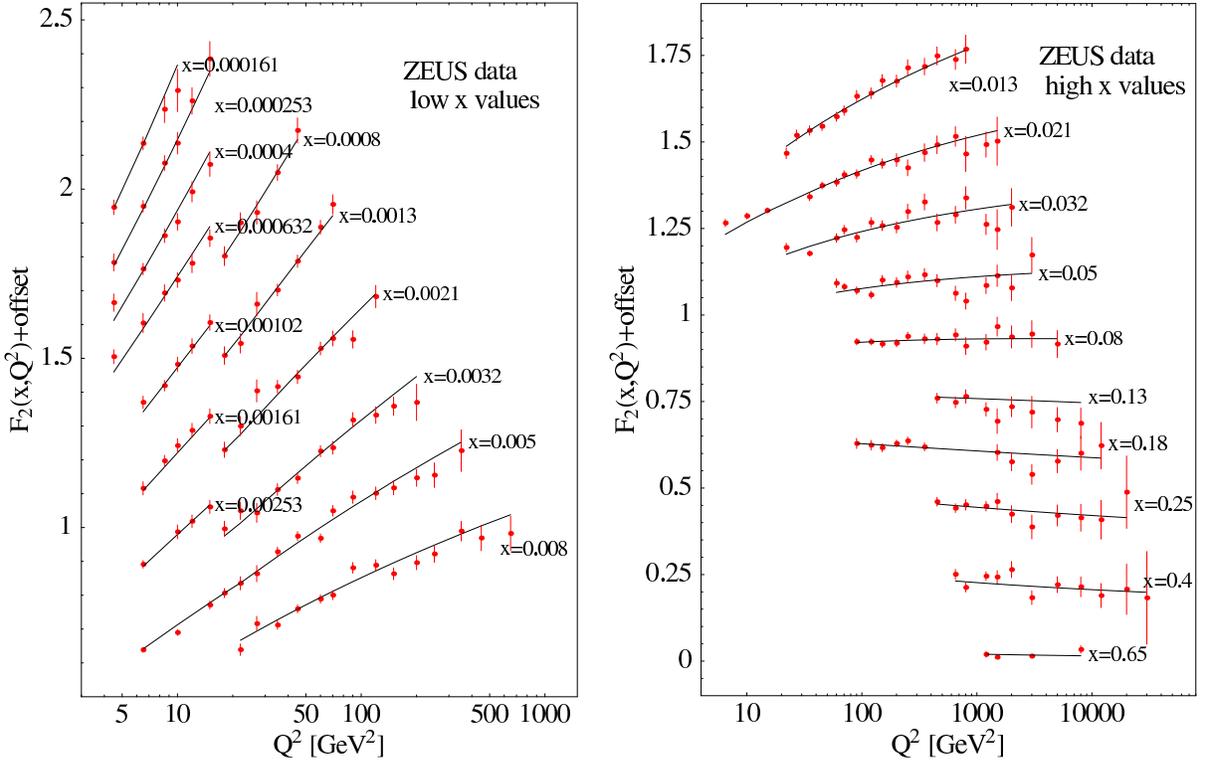
\figHone %
\figZEUS
The $\chi^2$
value is $263$ for $229$ data points. This is $2\sigma$
($\sigma=\sqrt{2N}=21$) away from the ideal value of $N=229$. A closer
inspection of Fig.~\ref{fig:ZEUS} does not suggest any systematic
disagreement. To assess the significance of this $2\sigma$ effect, we examine
in detail the systematic shifts obtained in the fit in
Appendix~\ref{app:DisErr}. We find that they are all quite reasonable, thus
giving us confidence that the fit is indeed of good quality.



The new PDF's also fit the older fixed-target DIS experiments well---similar
to previous global analyses. Figure \ref{fig:DisNcFt} shows the comparison to
the fixed-target neutral current experiments BCDMS and NMC. Because we are
incorporating the fully correlated systematic errors, the data sets used for
these experiments are those obtained at each measured incoming energy, rather
than the ``combined'' data sets that are usually shown.
This more detailed
and quantitative comparison is important when we try to evaluate the
statistical significance of the fits in our uncertainty analysis
(cf.\ Appendix~\ref{app:ErrorAna}).%
\figDisNcFt %
The $\chi^2$ per data point for these data sets are $1.11$ ($378/339$) for
BCDMS and $1.52$ ($305/201$) for NMC. The fit to the BCDMS data is clearly
excellent, both by inspection of Fig.~\ref{fig:DisNcFt}a and by the normal
$\chi^2$ test.  For the NMC data, Fig.~\ref{fig:DisNcFt}b shows rather good
overall agreement, but with some notable large fluctuations away from the
smooth theory curves. The most noticeable fluctuations---points with almost
the same $(x,Q)$ values---are from data sets taken at different incoming
energies.%
\footnote{\label{fn:Nmc}%
These fluctuations are smoothed out by re-binning and other measures in the
combined data set \cite{NMC}, which is not used here.
} %
This is
reflected in the $\chi^2$ value which is quite a bit larger than expected for
a normal probability distribution.
This raises two issues:
(i) Is the fit acceptable or unacceptable?
(ii) Can the fit be substantially improved by a different theoretical
model, e.g., the inclusion of higher-twist terms?
The first question is addressed in Sec.~\ref{app:Nmc},
where the excess $\chi^2$ is shown to be attributable to
larger-than-normal fluctuations of the data points
(which is unrelated to the viability of the theory model),
and that the shifts due to systematic errors
(which is related to the goodness-of-fit)
are all within the range of normal statistics.
The second question is addressed in Sec.~\ref{sec:HT} and
Appendix \ref{app:HT}, where we show that the inclusion of higher-twist
contributions does not result in a substantial improvement of the fit.

The charged-current DIS data from the CCFR experiment are
also used in our global analysis.
The agreement between data and theory for the CCFR measurements
of $F_{2}$ and $F_{3}$ is good, comparable to previous analyses.
Because of the important role of the charm quark in the
quantitative treatment of this process, and the lack of
information on correlations between the
current $F_2$ and $F_3$ data,%
\footnote{%
The CCFR $F_2$ measurements have been re-analyzed in a model-independent way
\cite{CCFR2}.
The new data are presented in a such a way that correlations
with the old $F_3$ measurements \cite{CCFR3} are no longer available.
\label{fn:ccfr}} %
we defer a detailed study of the charged-current DIS fit to the forthcoming
global analysis including heavy quark mass effects.

Of the new experimental input, perhaps the most interesting and significant
in its impact on the parton distributions is the one-jet inclusive cross
section from D\O\ \cite{D0jet}. These new data represent a considerable
expansion in kinematic range over the previous jet measurements, by providing
measurements in 5 separate rapidity bins, with $\eta$ up to $3.0$. Our fits
agree extremely well with these data, in both shape and magnitude over the
full kinematic range, as shown in Fig.~\ref{fig:IncJet}a.
 Comparison to
the corresponding CDF measurement in the central rapidity region is also
shown in the parallel plot, Fig.~\ref{fig:IncJet}b.%
\figIncJet%



The agreement between theory and experiment is made even clearer in
Fig.~\ref{fig:DzeroB} where the difference between data and theory is
plotted. The $\chi^2$ value for the fit to the jet data, including all
systematics, is $113/123$, combining the two experiments. In
Figs.~\ref{fig:IncJet} and \ref{fig:DzeroB}, the error bars are
combined statistical and diagonal systematic errors.%
\footnote{%
The D\O\ jet data involve partially correlated systematic errors which can
only be characterized by a full $N \times N$ covariance matrix $V$. Our
method of correcting for systematic errors on the data points, which depends
on writing $V$ in a separable form in terms of the $K\times N$ coefficients
\{$\beta_{ji}$\}, Eqs.~(\ref{eq:SysChi1}) and (\ref{eq:SysChi2}), cannot be
apply in this case.}
The ratio plot comparing to the CDF data is not shown; it is similar to
that of previous global fits, such as CTEQ5HJ. A detailed study of the impact
of these jet data on the determination of gluon distributions and the
potential for observing signals for new physics at the Tevatron and the LHC
will be carried out in a separate study.
\figDzeroB %

\subsubsection{The New Parton Distributions in the $\overline{MS}$ scheme}

\label{sec:C6mPdf}

Figure \ref{fig:CtqNewOld} shows an overview of the comparison between the
new PDF's and the previous generation of CTEQ PDF's, the CTEQ5M1 set, at
$Q=2$ GeV.%
\figCtqNewOld %
In order to exhibit the behavior of the PDF's clearly for both
large and small $x$ in one single plot, we choose the abscissa to be scaled
according to $x^{1/3}$. Correspondingly, we multiply the ordinate by the
factor $x^{5/3}$, so that the area under each curve is proportional to the
momentum fraction carried by that flavor in the relevant $x$ range. We see
that the most noticeable change occurs in the gluon distribution.

\paragraph{The gluon distribution}

Figure \ref{fig:GluCompare} gives a more detailed picture of the changes in
the gluon distribution at $Q=2$ and $100$\,GeV. For low and moderate values
of $x$, say $10^{-5}< x < 0.1$, the most important constraint is due to the
rate of $Q^2$-evolution of the DIS structure functions. The HERA data in this
region are ever improving in accuracy, but the new data has not made a
sizable change in the gluon distribution, as seen in
Figs.~\ref{fig:CtqNewOld} and \ref{fig:GluCompare}a. (Below $Q=2$ GeV, one
may find larger deviations between the new and old distributions, but
extrapolation of PDF's into that low-$Q$ region is well known to be unstable.
We will return to this point in Sec.~\ref{sec:Mrst01}.)%
\figGluCompare %


In the moderate to high $x$ range, $x > 0.01$, the inclusive jet data are now
playing a very important role. The combined effects of the precision DIS and
jet data have made a significant shift in the shape of the gluon
distribution, as seen in Figs.~\ref{fig:CtqNewOld} and \ref{fig:GluCompare}b.
The new gluon distribution is significantly harder than for CTEQ5M1 and all
MRST PDF sets (cf.\ Sec.~\ref{sec:Mrst01}) at all $Q$ scales. This behavior
is mainly dictated by the inclusive jet data, which lie in the range $50 < Q
< 500$\,GeV and $0.01 < x < 0.5$. The D\O\ data in the higher $\eta$ bins now
allow a higher $x$ reach than the central jet data from previous
measurements. The hard gluon distribution becomes amplified at lower $Q$
scales, due to the nature of QCD evolution and the fact that there is no
direct experimental handle on the gluon at large $x$ and low $Q$. The
enhanced gluon at large $x$ is similar to the CTEQ4HJ and CTEQ5HJ
distributions. However, there is an important difference in the significance
of the current result: whereas the ``HJ'' PDF sets were obtained specifically
for fitting the high $p_T$ jet data, by artificially inflating the weights of
those points in the global fit, the CTEQ6M gluon distribution results
naturally from the new global fit without any such special emphasis. The
visually good fits seen in Figs.~\ref{fig:IncJet} and \ref{fig:DzeroB} are
quantitatively substantiated by the small $\chi^2/N$ value of 113/123 for the
jet data sets. Since CTEQ6M represents the ``best fit'' in the global
analysis, this gluon behavior is also fully consistent with all DIS and
Drell-Yan data sets used in the fit, as discussed previously.

\subsubsection{DIS and LO Parton Distributions}

\label{sec:DisLo}

For most applications, the \mbox{\small {$\overline {MS}$}}\ parton
distributions are the most appropriate. But for certain applications, PDF's
in the NLO-DIS scheme are preferred. For these purposes we have obtained
CTEQ6D by performing independent global fits in the NLO-DIS scheme. Although,
in principle, one could obtain NLO-DIS parton distributions by a simple
transformation from a NLO-\mbox{\small {$\overline {MS}$}}\ set, the
reliability of such a procedure is uncertain in $x$ regions where the
numerical values of the PDF's for different flavors
(which transform into each other)
are orders of magnitude apart.%
\footnote{%
A recently discovered \cite{Dasgupta:2001eq} error in the QCD evolution of
DIS scheme PDF's in previous CTEQ analyses has been corrected in this work.
This error has negligible practical consequences in physical applications in
the current energy range, because the small deviation in evolution was
naturally compensated by the fitted PDF's. As long as the same DIS hard cross
sections are used in the fitting and in the applications, as is indeed the
case, the same physical cross sections will be reproduced. We thank G.\ Salam
for useful communications on this issue.} %
It is thus preferable to perform an independent fit. The quality of fit for
the NLO-DIS fit is comparable to that of the \mbox{\small {$\overline {MS}$}}
case.

Leading-order parton distributions are useful for Monte Carlo event generator
applications. For this purpose, we produce a set of LO PDF's, CTEQ6L.
These are obtained in a global fit with leading order hard cross sections.
In this fit, the NLO $\alpha_s$ (with $\alpha_s(m_Z)=0.118$)
is used for the QCD coupling, rather than the LO $\alpha_s$.
This combination of choices of LO hard cross sections
and NLO $\alpha_s$ conforms with the prevailing
practice in MC event generator applications.

\subsection{Eigenvector PDF Sets For Uncertainty Analyses}

\label{sec:PdfUnc}

As mentioned in the Introduction and in Sec.~\ref{sec:NewMethod}, an
important goal of this work is to advance global QCD analysis to include
systematic and quantitative estimates of uncertainties---on both PDF's and
their physical predictions. For this purpose, we characterize the behavior of
the global $\chi^2$ function in the neighborhood of the minimum by a set of
eigenvector PDF sets, according to the method of \cite{Hesse} %
(cf.\ the illustration in Sec.~\ref{sec:NewMethod}).

The eigenvector sets are obtained in two steps. First, the full set of
parameters described in Sec.~\ref{sec:Parametrization} is probed with the
iterative procedure of \cite{Multivariate,Hesse}, in order to identify those
parameters that are actually sensitive to the input data set.%
\footnote{%
The sensitive parameters are those that are not close to ``flat'' directions
in the overall parameter space, as mapped out by the iterative
diagonalization procedure of \cite{Multivariate}.} %
With current data, and our functional form, 20 such parameters are
identified. We then generate the eigenvector PDF sets in the 20 dimensional
parameter space as described in \cite{Multivariate,Hesse}, with the
remaining parameters held fixed. This results in 40 PDF sets, a $+$ (up) and
a $-$ (down) set for each eigenvector direction in the parameter space, in
addition to the central CTEQ6M set. Ideally, in the quadratic approximation
of the Hessian approach, the $\chi^2$ curves would be symmetric around the
minimum, so that only one displacement would be needed for each eigenvector.
However, we observe some asymmetry in certain directions in practice, so
we have decided to generate both up and down sets in each eigenvector
direction; this provides more information on the behavior
of constant-$\chi^2$ surfaces in the neighborhood of the minimum.%
\footnote{%
This asymmetry has been encountered in some applications, e.g.\ Ref.\cite{NadSul}.} %
The up/down sets, called $S_1^+, S_1^-,\dots,S_{20}^+,S_{20}^-$, correspond
to a \emph{tolerance} of $T=10$ (cf.\ Ref.~\cite{Hesse}); \i.e., their
$\chi^2$ value is greater than the minimum by $T^{2}=100$. The range of
uncertainty corresponding to the choice $T=10$ represents our estimate of the
range of PDF behavior that is consistent with the current data. Details of
the error analysis that leads to this estimate are described in Appendix\
\ref{app:tolerance}.

\subsubsection{Uncertainties of PDF's}

\label{sec:DelPdf}

We use the eigenvector PDF sets $\{S_{i}^{\pm}\}$ to estimate the uncertainty
range of physical quantities (and of the PDF's themselves) according to the
master equation (\ref{eq:DeltaX}) which is derived in \cite {Hesse}. In
Figs.~\ref{fig:QrkUnc} and \ref{fig:GluUnc} we show fractional uncertainty
bands for the $u$-quark, $d$-quark, and gluon distributions, respectively, at
$Q^2=10 \, \mathrm{GeV}^2$. In these plots, the shaded region is the envelope
of allowed variation of the parton distribution, independently for each value
of $x$, with tolerance $T=10$. The ordinate is the ratio of the extreme value
(up or down) to the standard CTEQ6M value. For comparison, the curves are
CTEQ5M1 (dashed) and MRST2001 (dotted), plotted in ratio to CTEQ6M.%
\figQrkUnc %

The $u$ distribution is the most accurately known of the parton
distributions, since deep-inelastic scattering by photon exchange, being
proportional to the square of the quark charge in leading order, is most
sensitive to the $u$ quark. The $d$-quark distribution is very much affected
by the various data sets that are sensitive to $u$-$d$ differences: the NC
and CC DIS measurements with proton and deuteron targets, the DY p/d
asymmetry and the W-lepton asymmetry experiments. The $d$-quark uncertainty
band is seen to be noticeably wider than that of the $u$ quark, particularly
at large $x$ where there are few constraints on the ratio between the two
flavors. This result provides quantitative confirmation of a previous study
\cite{CteqLgx}, addressing the issue of the behavior of the $d$-quark
distribution at large $ x$ raised in \cite{BodekYang}.

The gluon distribution is the most uncertain of the PDF's---notwithstanding
the increased constraints from recent precision DIS and inclusive jet data
discussed earlier---as shown in Fig.~\ref{fig:GluUnc}. The uncertainty is of
order $\pm{15}\%$ for $x$ values up to $\sim 0.3$, and then it increases
rapidly for large $x$. This uncertainty is of much interest for the physics
programs of the Tevatron and the LHC, and it has been the subject of several
previous studies \cite{PrevGluUnc}. With the new quantitative tools at hand,%
\figGluUnc %
we have found that it is important to use a sufficiently flexible
parametrization of the initial distribution, in order to obtain a reliable
estimate of the range of uncertainty of the gluon distribution. This
point will be relevant when we compare the above result with other recent
studies of the gluon distribution (cf.\ Sec.~\ref{sec:Mrst01}). Some details
are given in Appendix \ref{app:Param}.


The uncertainties shown in Figs.~\ref{fig:QrkUnc} and \ref{fig:GluUnc} are
significantly smaller than those obtained in our previous study
\cite{Hesse} based on CTEQ5 inputs.
The reduced uncertainty, which can be seen by comparing
Figs.~\ref{fig:QrkUnc} and \ref {fig:GluUnc} with their
counterparts (Figs.~3 and 4) of Ref.~\cite{Hesse},
is a consequence of the new DIS and jet data.%
\footnote{%
One must be aware that the uncertainties shown in \cite{Hesse} are for a
higher value of $Q=10 \, \mathrm{GeV}$. The uncertainties decrease when
evolved to higher $Q$.} %

An important point in the interpretation of
Figs.~\ref{fig:QrkUnc} and \ref{fig:GluUnc} is that the
uncertainty ranges correspond to the envelopes of possible
parton distributions that are consistent with the data.
A distribution function that produces the
extreme at any particular value of $x$ is generally not extreme at other
values of $x$. Thus, a PDF that follows the upper or lower boundary of the
uncertainty band at all $x$ would definitely not be consistent with the
data.

The CTEQ5M1 and MRST2001 distributions, plotted as ratios to CTEQ6M, are also
shown on Figs.\ \ref{fig:QrkUnc} and \ref{fig:GluUnc}, for comparison. Both
these PDF sets are generally within the current uncertainty bounds. The hard
gluon distribution in the CTEQ6M fit, discussed earlier, manifests itself in
the fact that both CTEQ5M1 and MRST2001 curves are close to the lower limit
of the uncertainty band at large $x$. For comparison, we have also included
the gluon distribution of CTEQ5HJ in Fig.\ \ref{fig:GluUnc}. It is even
harder than that of CTEQ6M at $x>0.3$, but still within the uncertainty band.
Comparisons of CTEQ6M and CTEQ5M1 fits are discussed in the next subsection
(\ref{sec:C5C6}). The MRST2001 distributions are rather similar to CTEQ5,
except at very small $x$ where the gluon is smaller and the quarks are
larger, both slightly outside the uncertainty ranges. Differences between
CTEQ6 and MRST2001 fits are discussed in some detail in
Sec.~\ref{sec:Mrst01}.

\subsection{Issues And Comments}

\label{sec:FitIssues}

\subsubsection{How much progress has been made?}

\label{sec:C5C6}


The fact that the new PDF's, especially the quark distributions, do not
appear to differ much from the CTEQ5 functions testifies to the steady
convergence of QCD analysis of PDF's. Progress in this undertaking must be
measured in more precise terms than previously, because of the increasing
demands for quantitative applications in precision SM studies and new physics
searches. To illustrate this point, we first compare the predictions of the
preceding generation of PDF's, CTEQ5M1, with two key new experiments, H1 and
D\O, in Figs.~\ref{fig:CfiveHone} and \ref{fig:CfiveDzero}. Plots like these
are often used in the literature as evidence of good or acceptable fits.
Careful comparison with the corresponding results for CTEQ6M
(Fig.~\ref{fig:Hone} and Fig.~\ref{fig:DzeroB}) reveals that the new fit is
discernably better.%
\figCfiveHone %

The improvement in the quality of the fits can be quantified by examining the
$ \chi^2$ values. With the inclusion of correlated systematic errors these
numbers now carry more statistical significance than before. The accompanying
table lists the CTEQ6M and CTEQ5M1 $\chi^2$'s for the full data sets used in
the current global analysis, and for the two representative experiments, H1
and D\O, shown in Figs.~\ref{fig:CfiveHone} and \ref{fig:CfiveDzero}.
The overall $\chi^2$ is greater by 268 (for 1811 data points) for the
previous generation of PDF's than the current one. The difference is not
very large,%
\footnote{%
In evaluating these $\chi^2$ numbers for the two fits, the overall
normalizations of all the data sets are allowed to re-adjust, within errors,
between the two calculations. Otherwise, the difference in $\chi^2$ would be
larger.} %
but it is outside our estimated tolerance of $T=10$ ($\Delta\chi^2=100$). %
 \figCfiveDzero %

The increases in the individual $\chi^2$'s for the two precision experiments,
although not very large, do lie outside the 90\% probability range according
to the uniform way we adopt for evaluating the uncertainty ranges,
discussed in Appendix \ref{app:tolerance}. These numbers provide a
quantitative measure of the improvement achieved. The precise interpretation
of these numbers is still open to discussion, however, and further progress
is still needed (cf.\ Appendix \ref{app:tolerance}).%
\begin{center}
\begin{tabular}{|c|c|c|c|}
\hline
$\chi^{2}$ & Total & H1 & D\O\ \\
(\# data pts.) & (1811) & (230) & (90) \\ \hline CTEQ6M & 1954 & 228 & 65 \\
\hline CTEQ5M1 & 2222 & 285 & 117 \\ \hline
\end{tabular}
\end{center}

\subsubsection{Higher twist terms?}
\label{sec:HT}

There have been several studies of higher-twist (HT) effects in DIS
experiments \cite{Botje,MRSThighertwist,AlekhinKataev}.
If power-law corrections to leading-twist PQCD are needed,
they will introduce additional nonperturbative degrees of freedom
in the global analysis.
This would complicate PDF analysis considerably,
because the extracted PDF's would then depend on the HT model.
Because there is no accepted theory of HT, and HT
terms are probably non-factorizable and process-dependent, PDF's obtained
with the inclusion of HT terms would no longer be universal.

In the absence of firm theoretical guidance, we first limit the possible
size of higher-twist terms by placing reasonable cuts on the kinematic range
of data utilized in the fitting program. We then study phenomenologically
the need for HT corrections by comparing the quality of fits with and
without HT parameters. Using the same kinematic cuts as in previous CTEQ
analyses (in particular, $Q \geq 2$\,GeV and $W \geq 3.5$\,GeV for DIS
data) we find that the inclusion of simple phenomenological HT factors, of
the type used in the previous literature
\cite{Botje,MRSThighertwist,AlekhinKataev},
does not produce discernable improvement in the quality of the fit.
We conclude that HT corrections are not needed,
and therefore stay within the twist-2 PQCD formalism for our
analysis.%
\footnote{%
This conclusion may appear to contradict some common lore about HT. We note
that most of the previous studies either were not done with as wide a range
of data sets, or did not incorporate the full correlated systematic errors.
Also, we use a relatively high $Q^2$ cut of $4\,\mathrm{GeV}^2$,
precisely to reduce the HT effects. } %
Our study of this issue is summarized in Appendix~\ref{app:HT}. We believe,
for the reasons mentioned in the previous paragraph, that should there be
evidence for HT effects, it would be more desirable to raise the kinematic
cuts and preserve the universal PDF's, rather than to introduce {\it ad hoc} HT
terms that would reduce the usefulness of the PDF's.

\subsubsection{Signs of anomalies at large $\mathbf{x}$ or small $\mathbf{x}$?}
\label{sec:Anomalies}

One of the important goals of quantitative global analysis of PDF's is to
provide stringent tests of the validity, and the efficacy, of PQCD. Anomalies
observed in one or more of the hard processes in the global analysis could
indicate signs of new physics.

Among the hard processes in current global analyses, the high-$E_{T}$ jet
cross section measured by CDF and D\O\ probes the smallest distance scales,
and hence provides the best window to discover new physics at large energy
scales. There was some excitement when CDF first measured a possible excess
in the high-$E_{T}$ jet cross section, compared to conventional PDF
predictions of that time \cite{Abe:1996wy}. This excitement has abated with
the advent of the ``PDF explanation'' \cite{Huston:1995tw} (the CTEQ4HJ and
CTEQ5HJ type of gluon distributions \cite{Cteq}) and the subsequent D\O\
measurements \cite{Abbott:1998ya} (which do not show as pronounced an
effect). However, the issue remains an interesting one since the CTEQ4HJ/5HJ
gluon distributions have not been universally accepted as the PDF of choice.
Where does the issue stand in view of the recent data and global analyses?

As already mentioned in Sec.~\ref{sec:C6mPdf} above, the CTEQ6M central fit
arrives at a gluon distribution that is considerably harder at high $x$ than
the conventional ones (such as CTEQ5M and MRST2001). It is more like CTEQ4HJ
and CTEQ5HJ, although obtained without giving any extra weight to the jet
data. This change comes about from three developments: (i) the statistical
power and the expanded range of the new D\O\ jet data set in the five
separate rapidity intervals \cite{D0jet}; (ii) the increased flexibility of
the nonperturbative gluon distribution in the new parametrization adopted for
this global analysis (cf.\ Secs.~\ref{sec:Parametrization}, \ref{sec:Mrst01}
and Appendix \ref{app:Param}); (iii) the fact that the harder gluon is needed
to provide a better fit to the D\O\ data. On one hand, this result provides
further support for the PDF explanation of the high-$E_{T}$ jet cross
section. On the other hand, by the more quantitative uncertainty analysis
shown in Fig.~\ref{fig:GluUnc}, there is still a large range of possible
behavior for the high-$x$ gluon, allowing both the conventional and the
``HJ-like'' shapes. Further work will be required to draw a firm conclusion
on this important issue. Continued search for signs of new physics must be
pursued with vigor in all channels.

Signals for a departure from conventional PQCD could also appear in the very
small $x$ region. Our analysis is based on the standard NLO DGLAP formalism.
The impressive agreement between the fits and all the data sets shows no
indication of a breakdown of conventional PQCD; nor do we detect any
necessity for including NNLO corrections \cite{AlekhinNnlo}.  This does not
mean that effects beyond the simple theoretical model are not present.  It
does mean that the search for these effects must depend on even more precise,
and extensive, experimental and theoretical input.

\subsubsection{Determination of $\protect\alpha_s$?}

\label{sec:alpha}

As mentioned in Sec.~\ref{sec:Parametrization}, for our standard analysis we
input the strong-coupling strength $\alpha_s(m_Z) = 0.118$, based on
dedicated measurements from QCD studies at $e^+e^-$ colliders and sum rules
in lepton-hadron processes. It is desirable to check that this value is
consistent with our global analysis, and, beyond that, to see whether the
global analysis can provide a useful independent measurement of $\alpha_s$.

For this purpose, we have repeated the fitting with different choices of
$\alpha_s(m_Z)$. The resulting variation of the global $\chi^2$ is shown in
Fig.~\ref{fig:AlfChi}. The minimum in $\chi^2$ occurs at a value of
$\alpha_s(m_Z) \simeq 0.1165$, somewhat lower than the world average from
precision measurements. For our choice of $\alpha_{s}(m_{Z})=0.118$, $\chi^2$
is greater than the minimum by about $5$. This difference is completely
insignificant with respect to our estimated tolerance of
$\Delta\chi^{2}<T^{2}\approx 100$. Thus, the value of $\alpha_s(m_Z)$ favored
by global PDF analysis is in excellent agreement with the world average.
Furthermore, according to Fig.~\ref{fig:AlfChi} and our adopted tolerance,
the allowed range of $\alpha_s(m_Z)$, as measured in global PDF analysis, is
about 0.110 to 0.123 (i.e.\ $\alpha_s(m_Z)=0.1165 \pm 0.0065)$. This is not
competitive with other dedicated measurements. The basic reason is
well known: the uncertainty of $\alpha_s$ determination in QCD analysis of
parton distributions is strongly tied to the uncertainty of the gluon
distribution, which still has substantial uncertainties.%
\figAlfChi


We also note that the definition of NLO $\alpha_s(\mu)$ in perturbative QCD
is subject to an ambiguity relating to the solution of the renormalization
group equation (RGE). We have used the conventional \mbox{\small {$\overline
{MS}$}}\ formula for $\alpha_{s}(\mu)$, obtained by solving the NLO RGE as an
expansion in inverse powers of $\ln \mu$ \cite{PDG}.
An alternative
definition, favored by some, and more convenient for higher-order extensions,
is to numerically solve the truncated RGE equation for $\mu \, d\alpha/d\mu$
at NLO. The dotted curve in Fig.~\ref{fig:AlfChi} shows $\chi^{2}$ versus
$\alpha_{s}(m_{Z})$ using this alternative definition of $\alpha_{s}(\mu)$.
The difference between the two treatments is seen to be within the range of
uncertainties.

\subsection{User Interface}

\label{sec:Interface}

All PDF sets described above will be available (at http://cteq.org) in the
usual CTEQ format, using external data tables. These can be used in the same
way as previous CTEQ PDF's.

In addition, to facilitate progress toward a universal user interface, we are
preparing a new format, which uses an evolution program and small external
files containing only the coefficients for the initial parton distributions,
following the ``Les Houches Accord on PDF's'' \cite{LHaccord}.

\section{Physical Predictions and their Uncertainties}

\label{sec:PhyPred}

The main utility of the new parton distribution functions is to make
predictions on physical cross sections and their uncertainty ranges---both
for precision SM studies and for new physics searches. Detailed applications
go beyond the scope of this paper. We will, however, present some general
results on estimated uncertainties of various parton luminosity functions for
the Tevatron (RunII) and the LHC. From these results, one can easily estimate
the expected uncertainties of a variety of physical processes of interest for
these hadron collider programs: W/Z cross sections, Higgs, and top (both
single and pair production) rates due to various production mechanisms, etc.


Figure \ref{fig:LumiTeV}a shows the fractional uncertainties of the
gluon-gluon and quark-antiquark luminosity functions, to final states with
the quantum numbers of $W^\pm$, $W^-$, $\gamma^*$ and $Z^0$, at the Tevatron.%
\figLumiTeV %
Figure \ref {fig:LumiTeV}b similarly displays the quark-gluon
luminosity functions for $W^+,W^-,\gamma^*,Z$ final-state quantum numbers.
As expected, the gluon-gluon luminosity has the greatest uncertainty, %
especially at large $\sqrt{\hat{s}}$, reflecting the uncertainty of the gluon
distribution at large $x$. The $q\bar{q} \rightarrow W^+,W^-,\gamma^*,Z$
luminosity uncertainties are almost constant throughout the range of
$\sqrt{\hat{s}}$ plotted, at the $\pm 4\%$ level. The $qG \rightarrow
W^+,W^-,\gamma^*,Z$ luminosity uncertainties resemble the geometric mean
between those for $GG$ and $qG$. The four $qG$ uncertainty bands are very
similar to one another, but close examination shows that they are not
identical, because of the differences in quark flavor mix.

Similarly, Fig.~\ref{fig:LumiLhc} shows the fractional uncertainties of the
corresponding luminosity functions for the LHC. The $q\bar{q} \rightarrow
W^\pm$ luminosity uncertainties are also fairly constant,
being $\pm 5\% $ for $q\bar{q} \rightarrow W^+$ and $\pm 4\%$ for $%
q\bar{q} \rightarrow W^-$ for $100 < \sqrt{\hat{s}} < 200 \:\mathrm{GeV}$.%
\figLumiLhc
The gluon-gluon luminosity uncertainty is $\pm 3\% $ at the narrowest point
($\sqrt{\hat{s}} \sim 250\; \mathrm{GeV})$, increasing to $\pm 10\;\%$ at
both ends of the $\sqrt{\hat{s}}$ range shown. From this, one can estimate
that the uncertainty on the production cross section of a light mass Higgs
particle (say, $100 \mathrm{-} 200$\,GeV) at the LHC due to PDF's
is on the order of $\pm 5\%$. The uncertainty of the gluon-gluon luminosity
function at the LHC is seen to be generally smaller than for the Tevatron
case. The difference results from the combination of two effects: the
uncertainty decreases with evolution, and the contributing $x$ ranges for the
two cases are somewhat different. The luminosity uncertainties for the
Compton processes at the LHC, Fig.~\ref {fig:LumiLhc}b, are again rather
similar to one another. The uncertainties
range from around $\pm 6\%$ at low energies to $\pm 2\;\mathrm{-}\;3\%$ at $%
\sqrt{\hat{s}}\sim 500\;\mathrm{GeV}$.


The W cross section can be measured with great precision at hadron-hadron
colliders. The cross section is large and the backgrounds are relatively
small. In addition, the theoretical uncertainties are small. This makes the W
cross section an ideal benchmark with which to normalize other cross
sections, especially as there remains a significant uncertainty as to the
value of the total inelastic cross section at the Tevatron. (The inelastic
cross section is used by the experimental collaborations to normalize all
luminosity calculations. CDF and D\O\ currently assume values for this cross
section that differ by 6\%, producing cross-section differences between the
two experiments of the same amount.)  The simple estimates of the $W$ cross
section uncertainty the PDF uncertainties at the Tevatron and the LHC, given
above, can be improved by including NLO terms and by using the more precise
Lagrange method \cite{Lagrange}. Using the 40 Hessian eigenvector sets, the
uncertainty for any cross section can be calculated. This procedure will be
made more straightforward with the adoption of the Les Houches accord on
PDF's.

\section{Comparison with Other Parton Distribution Analyses}

\label{sec:Compare}

\subsection{Previous Uncertainty Studies}

The quantitative analysis of uncertainties of PDF's, taking into account the
experimental systematic errors, was pioneered by
\cite{Alekhin,ZomerDis,Botje}, in the context of detailed study of the
precision DIS experiments. These studies were based on $\chi^2$ minimization
techniques. The methods that we developed in
\cite{Multivariate,Hesse,Lagrange}, and used in this paper, have extended
this approach to a global analysis (cf.\ the recent review \cite{BotjeRev}.)
Because the scope of the experimental data used in the earlier studies is
different from ours, direct comparison with those results is not possible.

Giele et al.\ (GKK) advocate a more general approach, using a pure
probabilistic analysis with random sampling of the parton parameter space
\cite{GKK}. While theoretically appealing, the proposed method encounters
difficulties when confronted with real experimental data sets, which either
do not meet the strict statistical criteria or are mutually incompatible
according to a rigorous probabilistic interpretation of the errors. As a
result, although the methodology is clearly established, it has not yet
produced practical results that can be generally used.

In an ideal world, where all experimental data sets are self-consistent and
mutually compatible, the GKK method is equivalent to the one we use,
provided, in addition, the Gaussian approximation underlying the $\chi^2$
approach is valid. If, however, $\chi^2$ is not approximately quadratic in
the neighborhood of the minimum, and if the precise response functions for
all experiments in the global analysis are known, then the GKK method would
be superior. Minimization of $\chi^{2}$ could still be useful in that case,
but the accuracy is only as good as the Gaussian approximation. In reality,
of course, the detailed experimental response functions are rarely, if ever,
known. So realistically the study of PDF uncertainty will necessarily consist
of quantitative statistical analysis supplemented by pragmatic choices based
on physics considerations and experience.

\subsection{Comparison to MRST2001}

\label{sec:Mrst01}

Our results can be compared more directly to those of the recent MRST2001
analysis \cite{Mrst01}. An overview of the comparison between the CTEQ6M and
MRST2001 parton distributions is shown in Figure \ref {fig:CsixMrst}.
Significant differences appear in the gluon distribution, similar to the
differences between CTEQ6 and CTEQ5 described in Sec.~\ref {sec:C6mPdf}.
Unlike CTEQ5, the MRST analysis uses essentially the same recent experimental
input as CTEQ6. Thus it is natural to seek the sources of the observed
differences.


Although most of the experiments used in the CTEQ6 and MRST2001 analyses are
the same, there are some differences both in data sets and in the way that
experimental error information is utilized. The CTEQ6 analysis uses the
separate-energy data sets of the BCDMS and NMC experiments (for which
information on correlated systematic errors is available) whereas MRST uses
the combined-energy sets for these experiments. The CTEQ6 analysis does not
use SLAC experiments, because that data lies in the region of very low $Q^2$.
Also, CTEQ6 does not use the HERA charm-production data, because the errors
are still large, and because a proper theoretical treatment of this process
requires the full inclusion of charm mass effects in the matrix
elements.\footnote{%
Work is under way for inclusion of charm production in a separate global
analysis.} In addition, the kinematic cuts on data selection are somewhat
different.%
\figCsixMrst %

There are several differences in methodology. In the analysis of data, the
published experimental correlated systematic errors are fully incorporated in
the CTEQ6 global fitting procedure. In the theoretical calculations, we use
the conventional (zero-mass parton) Wilson coefficients for DIS structure
functions for reasons explained in Sec.~\ref{sec:QcdParam}, whereas MRST uses
the Thorne-Roberts version of the VFNS calculation \cite{TR}. Finally, the
two groups use different functional forms for the parametrization of the
nonperturbative input parton distributions.

Beyond a visual inspection of the PDF's, insight can be gained by comparing
the two fits for data sets used in both analyses. In principle, one might
just compare the $\chi^2$ values of the corresponding experiments for the two
fits, cf.\ $\chi^2$ numbers given in Appendix \ref{app:C6m} and Table 1 of
Ref.~\cite{Mrst01}. In practice, a direct comparison between the DIS
experiments is not possible because the two analyses have different
treatments of the experimental systematic errors and different choices of
Wilson coefficients mentioned above. Among experiments that can be directly
compared, two differences are noticeable. The first experiment is E605
(Drell-Yan): $\chi^2/N$ is given as $232/136$ in MRST2001 \cite{Mrst01},
and $95/119$ for CTEQ6M.%
\footnote{%
The differences in the number of data points are due to slightly different
kinematic cuts in the two analyses.} %
However, the seemingly large MRST number was due to their use of statistical
errors only for this experiment. The two fits are actually comparable for
this experiment.  The second is the combined D\O\ and CDF jet cross section
measurements: $\chi^2/N$ is $170/113$ for MRST2001, and $113/123$ for CTEQ6M.
This difference is statistically quite significant, and it is responsible for
the rather distinct gluon distributions obtained by the two groups.  We
believe that this difference is primarily due to the parametrization of the
nonperturbative parton distributions, especially the gluon, as will be
discussed below and in Appendix \ref{app:Param}.

\vspace{2ex}
\noindent We now comment more specifically on a number of issues highlighted
in the MRST paper \cite{Mrst01} from the perspective of the current study.
\vspace{-2ex}

\paragraph{Large-$\mathbf{x}$ gluon behavior}

The inclusive jet data have a strong influence in the determination of the
gluon distribution at medium and large $x$. The MRST analysis found a rather
significant competition between the jet data and the Drell-Yan (E605) data in
the context of their global fit. This requires either a compromise, the
MSRT2001 set, or a good fit to the jet data with a much poorer Drell-Yan fit
(and a rather unusual gluon shape), the MRST2001J set \cite{Mrst01}.

This dilemma is not observed in our analysis. The CTEQ6M fit to both the jet
data and the Drell-Yan data appear nearly ``ideal'' (i.e., $\chi^{2}/N \sim 1
$), and the gluon shape is rather smooth. This clear difference results
mainly from the parametrization of the PDF's, especially that of the gluon.
We are led to this conclusion by a study where we adopt the MRST
parametrization for the gluon and repeat the global fit. We then obtain a
``best fit'' in which the gluon has a shape similar to MRST2001J. The gluon
distributions at $Q=2$\,GeV for CTEQ6M, MRST2001, MRST2001J, and the
MRST-like fit that we obtain by adopting the MRST parametrization are
compared in Figure \ref {fig:MrstGlu}. The ``wiggle'' of the gluon function
observed in the latter two fits (which led MRST to reject the MRST2001J fit
\cite{Mrst01}) is clearly due to the particular functional form. More
details, including other examples and results at $Q=1\;\mathrm{GeV}$ (which
are more sensitive to the differences), are presented in Appendix
\ref{app:Param} on the question of parametrization.
\figMrstGlu %


\paragraph{Gluon behavior at small $\mathbf{x}$ and small $\mathbf{Q}$}

The MRST study found a strong preference for a negative gluon distribution at
small $x$ and small $Q$. It has been known for a long time that QCD evolution
is extremely rapid at small $Q$. This makes extrapolation \emph{backward} to
$Q=1 \, \mathrm{GeV}$ quite unstable \cite{JCC86}, even though the behavior
of the PDF's at $Q > 2 \, \mathrm{GeV}$ is rather tightly constrained by the
experimental data. It is generally agreed that there is no theoretical
requirement that parton distributions be positive definite at any particular
$Q$, as long as cross section predictions stay positive. The issue is, thus,
only a phenomenological one.

We have chosen $Q_0 = 1.3  \, \mathrm{GeV}$ for the CTEQ6 fit, and find no
indication for $G(x,Q)$ going negative in the region $Q > 1.3  \,
\mathrm{GeV}$.  To consider lower values of $Q$, we have made alternative
fits with $Q_0$ down to $1.0  \, \mathrm{GeV}$ which are described in
Appendix \ref{app:Param}. The results can be summarized as follows: (i)
Whether or not the gluon distribution tends to go negative is strongly
dependent on the choice of $Q_0$---it will always go negative at some low
$Q$, as anticipated in \cite{JCC86}. (ii) With $Q_0 = 1\;\mathrm{GeV}$,
parametrizations that allow for a negative gluon at small $x$ can yield
slightly lower $\chi^2$ than those with positive-definite gluons. This has no
phenomenological significance, however, since the difference between the
resulting gluon distributions is much smaller than its uncertainty in this
region. The uncertainty assessment is closely related to the study of
parametrization described in Appendix \ref{app:Param}.  Note, in particular,
Fig.~\ref{fig:NegFit}b.

\paragraph{Measurement of $\mathbf{\protect\alpha_{s}}$}
MRST performed an analysis to determine $\alpha_s(m_Z)$ from their global
fits, obtaining $\alpha_S(m_Z) = 0.119 \pm 0.002 \, \hbox{(\rm expt.)} \pm
0.003 \,\hbox{(\rm theory)}$. As discussed in Sec.~\ref{sec:alpha}, our study
obtained a lower central value and a substantially larger error estimate. The
latter ($\pm 0.002$ vs.\ $\pm 0.0065$) is mainly due to the different choices
of allowed $\Delta\chi^2_{\mathrm{tot}}$ (20 vs.\ 100). In fact, in order to
achieve $\Delta\alpha_S(m_Z) = \pm 0.002\;\mathrm{(expt.)}$ from Fig.\
\ref{fig:AlfChi}, we would have to use $\Delta\chi^2_{\mathrm{tot}} \sim 10$.
In view of our study of the overall tolerance parameter discussed in Sec.\
\ref{app:tolerance}, we cannot justify having such a narrow range for
$\alpha_s(m_Z)$, since many perfectly acceptable global fits exist within the
full range $\alpha_S(m_Z) = 0.1165 \pm 0.0065$, even without including
theoretical uncertainties.

\section{Concluding Remarks}
\label{sec:Conclusion}

In this paper, a new generation of parton distribution functions has been
presented, making full use of the constraints of both the old and new data.
This global analysis significantly expands the scope of the currently used
PDF analyses on two fronts: (i) a full treatment of available experimental
correlated systematic errors, and (ii) a systematic treatment of
uncertainties of the parton distributions and their physical predictions,
using a recently developed eigenvector-basis approach to the Hessian method.
Thus, in addition to obtaining new best-fit PDF's, the ranges of
uncertainties of the PDF's and their physical predictions are systematically
assessed.

Among the improvements made in PDF determination, progress on the gluon
distribution is particularly worth noting. The new D\O\ jet data play a
significant role in obtaining a hard gluon distribution at large $x$---quite
different from previous standard fits.  We have shown that, at $Q^2 = 10
\:\mathrm{GeV^2}$, its uncertainty has been narrowed by recent precision data
to $\pm 10\%$ at small $x$ (up to $x \simeq 0.3$). The uncertainty then
increases quite rapidly (in spite of the improved constraints), reaching a
factor of 2 by $x \sim 0.5$. The need for additional precise data to pin down
the gluon distribution over a larger $x$ range is obvious.  The excellent
agreement between the new D\O\ non-central jet cross sections and the
precision DIS and DY data sets, within the PQCD framework using the new CTEQ6
PDF's, is significant. It lends credence to the PDF explanation of the
earlier CDF high $p_T$ jet measurement. A crucial element in achieving this
consistent picture is to allow sufficient flexibility in parametrizing  the
nonperturbative parton distributions, in order to accommodate a range of
possible behaviors at both low and high $x$.

The eigenvector basis PDF's obtained by this analysis allow the calculation
of the uncertainty of any physical quantity that depends on the PDF's.  We
have presented results on parton-parton luminosity function uncertainties for
both the Tevatron and the LHC. From these, uncertainties of a wide range of
SM and new physics signals and backgrounds can be inferred.  For instance,
the uncertainties on the production cross sections of $W,Z$ at the Tevatron
and the LHC are estimated to be $\pm 4\%$ and $\pm 5\%$ respectively, and
that of a light Higgs at the LHC to be $\pm 5\%$.  More detailed calculations
on the predicted cross sections and their uncertainties for processes of
interest can easily be done, using the standard CTEQ6M and its associated
eigenvector basis PDF's, or using the Lagrange method described in
Ref.~\cite{Lagrange}.

This work marks only the first step in the effort to advance the global
analysis of PDF's, incorporating the systematic treatment of uncertainties.
Improvements and applications of the results discussed in this paper are
worth pursuing.  It is also important to advance the theoretical input to the
global analysis, including resummation effects, improved treatment of heavy
quarks, and higher-order corrections. The current analysis is carried out at
NLO and achieves a satisfactory fit to all of the data sets over a wide
kinematic range. In the future, an extension to NNLO will be possible but
this must wait until more cross-section calculations are available at the
higher order, and until the precision of the experimental measurements makes
it useful. In the near future, new data from both the Tevatron and HERA,
along with further theoretical development, will allow for even more precise
determinations of PDF's, leading to more useful predictions for LHC physics.

\paragraph{Acknowledgements}
We are pleased to acknowledge fruitful discussions and collaborations with
many CTEQ colleagues over the years. Valuable suggestions on improvements to
this manuscript have been provided by F.\ Olness, J.\ Owens, and D. Soper.
J.\,P.\ and J.\,H.\ would like to thank many participants of the 2001 Les
Houches Workshop, particularly S.\ Alekhin, M.\ Botje, W.\ Giele, and A.\
Vogt, for very active interaction on general issues relating to global QCD
analysis of parton distributions.  WKT acknowledges several stimulating
discussions with M.\ Botje, W.\ Giele, and F.\ Zomer on the same topics.

The research of the authors is supported by the US Department of Energy, the
National Science Foundation, the Lightner-Sams foundation, and the Taiwan
National Science Council.

                           \clearpage
\appendix
\section{Details of the CTEQ6M fit}
\label{app:C6m}

The coefficients for the nonperturbative PDF's at $Q_0=1.3\:\mathrm{GeV}$, as
defined in Sec.\ \ref{sec:Parametrization}, for the CTEQ6M fit are:

\begin{center}
\begin{tabular}{|r|rrrrrr|}
 \hline
 & $A_0 \;$  & $A_1 \; \rule{0em}{3ex}$
 & $A_2 \;$  & $A_3 \; \rule{0em}{3ex}$
 & $A_4 \;$  & $A_5 \; \rule{0em}{3ex}$ \\
 \hline
$ \rule{0em}{3ex} d_v$                & $ 1.4473$ & $ 0.6160$ & $4.9670$
& $-0.8408$ & $0.4031$ & $ 3.0000$ \\
$ \rule{0em}{3ex} u_v$                & $ 1.7199$ & $ 0.5526$ & $2.9009$
& $-2.3502$ & $1.6123$ & $ 1.5917$ \\
$   \rule{0em}{3ex} g$                & $30.4571$ & $ 0.5100$ & $2.3823$
& $ 4.3945$ & $2.3550$ & $-3.0000$ \\
$ \rule{0em}{3ex} \bar{u}+\bar{d}$    & $ 0.0616$ & $-0.2990$ & $7.7170$
& $-0.5283$ & $4.7539$ & $ 0.6137$ \\
$ \rule{0em}{3ex} s = \bar{s}$        & $ 0.0123$ & $-0.2990$ & $7.7170$
& $-0.5283$ & $4.7539$ & $ 0.6137$ \\
$ \rule{0em}{3ex} \bar{d} / \bar{u}$  & $33657.8$ & $4.2767$
& $14.8586$ & $ 17.0000$ & $ 8.6408$ &  --\rule{1em}{0ex} \\
 \hline
\end{tabular}

\end{center}

\noindent The best fit to data is obtained with the following normalization
factors for the experiments:%
\footnote{\label{fn:H1}%
H1a and H1b refer to H1 data sets for low-$Q$ and high-$Q$ measurements
respectively \cite{H1}.}
\begin{center}
\begin{tabular}{|c|ccccccccc|}
\hline
 Expt.($e$)& BCDMS& H1a & H1b  & ZEUS & NMC  & CCFR & E605 & D\O\   & CDF \rule{0em}{3ex}\\
 \hline
 Norm($e$)& 0.976& 1.010& 0.988& 0.997& 1.011& 1.020& 0.950& 0.974& 1.004 \rule{0em}{3ex}\\
 \hline
\end{tabular}
\end{center}

\noindent Table \ref{tbl:ChiAll} shows the $\chi^{2}$ values for those
experimental data sets for which detailed information on systematic errors is
available and used in the CTEQ6M fit. For each data set ($e$), we show: the number of data points ($N_e$), the $\chi^2$ value for that
experiment in the CTEQ6M fit ($\chi^{2}_{e}$), and $\chi^{2}_{e}/N_e$. %
\vspace{-4ex}%
\begin{table}[h]
\centering
\caption{Comparison of the CTEQ6M fit to data with correlated
systematic errors. \label{tbl:ChiAll}} \vspace{2.ex}
\begin{tabular}{|l|rrr|}
\hline data set & $N_e$ & $\chi^{2}_{e}$ & $\chi^{2}_{e}/N_e$ \rule{0em}{3ex}
\\
\hline BCDMS p & 339 & 377.6 & 1.114 \rule{0em}{3ex}
\\
BCDMS d & 251 & 279.7 & 1.114 \rule{0em}{3ex}
\\
H1a & 104 & 98.59 & 0.948 \rule{0em}{3ex}
\\
H1b & 126 & 129.1 & 1.024 \rule{0em}{3ex}
\\
ZEUS & 229 & 262.6 & 1.147 \rule{0em}{3ex}
\\
NMC F2p & 201 & 304.9 & 1.517 \rule{0em}{3ex}
\\
NMC F2d/p & 123 & 111.8 & 0.909 \rule{0em}{3ex}
\\
D\O\ jet & 90 & 64.86 & 0.721 \rule{0em}{3ex}
\\
CDF jet & 33 & 48.57 & 1.472 \rule{0em}{3ex}
\\
\hline
\end{tabular}
\end{table}

\noindent These results form the
basis for much of the quantitative uncertainty analysis discussed in this
paper. The interpretation of the ZEUS and NMC $\chi^2$'s are studied in some
detail in Sec.\ \ref{app:DisErr}.%
\footnote{%
The fairly large value of $\chi^2_e/N_e$ for the CDF data has been studied
extensively by the experimental group \cite{CDFjet}.  A substantial part of
the excess $\chi^2$ is due to fluctuations of a few points at moderate $p_T$.
The best value of $\chi^2_e/N_e$ that can be achieved for this data set
(using a high order polynomial curve fit to the data) is approximately
1.3. } %

Data sets with only effective uncorrelated errors are CCFR,%
\footnote{See footnote \ref{fn:ccfr}, Sec.\ \ref{sec:CompData}.} %
E605, E866, and CDF W-lepton asymmetry. The nominal $\chi^2_e\:/\: N_e$ for
these data sets for the CTEQ6M fit are (150/156, 95/119, 6/15, 10/11)
respectively.

\section{Comparison of Theory and Data with Correlated Errors}

\label{app:ErrorAna}

This Appendix consists of (i) a brief summary of the formalism used in our
error analysis---both for the $\chi^2$ calculations in performing the global
fits and for the interpretation of the results of those fits; (ii) an
examination of the significance of the observed higher-than-nominal $\chi^2$%
's for two of the DIS experiments; (iii) a discussion of the estimated
tolerance for our uncertainty estimates. Further details of the formalism can
be found in Ref.~\cite{Lagrange}.

\subsection{Useful formulas for $\protect\chi^2$ and the
analysis of systematic errors}

\label{app:SysErr}

The simplest $\chi^{2}$ function, used in most conventional PDF analyses, is
\begin{equation}
\chi_{0}^{2}=\sum_{\mathrm{expt.}}\ \sum_{i=1}^{N_{e}}
\frac{\left(D_{i}-T_{i}%
\right)^{2}}{\sigma^{\prime{2}}_{i}}  \label{eq:Chi2s}
\end{equation}
where $D_{i}$ is a data value, $T_{i}$ is the corresponding theory value
(which depends on the PDF model parameters {$\{a\}$), and
$\sigma^{\prime}_{i}$ is the combined statistical and systematic errors
(assumed uncorrelated and usually added in quadrature) on the measurement
$D_{i}$. This effective $\chi^{2}$ function provides a simple measure of
goodness-of-fit, convenient for the search for candidate PDF sets by
minimization. However, it is not useful for estimating the uncertainties
associated with those candidates because it does not contain enough
information to allow a meaningful statistical inference based on the increase
in $\chi^2$ away from the minimum.

Most DIS experiments now provide more detailed information on
measurement errors. For each data point $i$, we have the statistical error $%
\sigma_{i}$, uncorrelated systematic error $u_{i}$, and several (say, $K$)
sources of correlated systematic errors $\{\beta_{1i}, \beta_{2i}, \dots,
\beta_{Ki}\}$. The best fit to the data (i.e., the fit with least variance)
comes from minimizing the $\chi^{2}$ function,%
\footnote{%
The prime is to distinguish it from the its simplified---but
equivalent---form, $\chi^2$, derived below in Eq.\ \ref{eq:chi2corr}.}
\begin{equation}
\chi'^{2} (\{a\},\{r\})= \sum_{\mathrm{expt.}}\ \left[ \sum_{i=1}^{N_{e}} \frac{1}{%
\alpha_{i}^{2}} \left(D_{i}-T_{i}-\sum_{k=1}^{K}r_{k}\beta_{ki}\right)^{2}
+\sum_{k=1}^{K} r_{k}^{2}\right] \label{eq:Chi2sys}
\end{equation}
where $\alpha_{i}^{2}=\sigma_{i}^{2}+u_{i}^{2}$ is the combined uncorrelated
error. The fitting parameters are (i) the PDF model parameters $\{a\}$, on which $%
T_{i}$ depends, together with (ii) random parameters $\{r\}$ associated with
the sources of correlated systematic error. The
point of Eq.~(\ref{eq:Chi2sys}) is that $D_{i}$ has a fluctuation $%
\sum_{k}r_{k}\beta_{ki}$ due to systematics. The best estimate of this shift
is obtained by minimizing $\chi'^{2}$ with respect to the set $\{r_{k}\}$. In
practice, the total number of such parameters for all experiments included in
the global analysis can be quite large. Adding these to the already large
number of PDF parameters $\{a\}$ (which represent the real goal of the
analysis) one encounters a formidable minimization task, involving a
parameter space of dimension close to 100. The practical difficulties have
considerably hindered past efforts using this approach. The stability and
reliability of results obtained this way can also be questioned.

We pointed out in Ref.~\cite{Lagrange} that the minimization of the function
$\chi^{2}$ with respect to $\{r\}$ can be carried out analytically. This
simplifies the global analysis to its irreducible task of minimization with
respect to the PDF parameters $\{a\}$ only.  In addition, the analytic method
provides explicit formulas for the optimal values of $\{r_{k},k=1\dots K\}$
due to the systematic errors $k=1\dots{K}$ that are associated with the fit.
These optimal shifts are obtained from the condition
$\partial\chi^{2}/\partial{r}_{k}=0$, and the result is
\begin{equation}
r_{k}(\{a\})=\sum_{k^{\prime}=1}^{K}\left(A^{-1}\right)_{kk^{\prime}}\,B_{k^{%
\prime}}\;\; .  \label{eq:r=}
\end{equation}
Here $\{B_{k}\}$  and $\{A_{kk^{\prime}}\}$ are given by
\begin{equation}
B_{k}(\{a\})=\sum_{i=1}^{N_{e}}\frac{\beta _{ki}\left( D_{i}-T_{i}\right)
}{\alpha _{i}^{2}} ~~~~ {\rm and} ~~~~
A_{kk'}=\delta _{kk^{\prime }}+\sum_{i=1}^{N_{e}}%
\frac{\beta _{ki}\beta _{k^{\prime }i}}{\alpha _{i}^{2}} \;\:.
\label{eq:BandA}
\end{equation}
Substituting the best estimates (\ref{eq:r=}) back into $\chi'^{2}$, we obtain
a simplified $\chi^{2}$ function,
\begin{equation}
\chi^{2}(\{a\}) \equiv \chi'^{2} (\{a\},\{r(\{a\})\})
=\sum_{\mathrm{expt.}}\ \left\{ \sum_{i=1}^{N_{e}} \frac{%
\left(D_{i}-T_{i}\right)^{2}}{\alpha_{i}^{2}} -\sum_{k,k^{\prime}=1}^{K}
B_{k}\left(A^{-1}\right)_{kk^{\prime}} B_{k^{\prime}} \right\}.
\label{eq:chi2corr}
\end{equation}
Minimizing $\chi^{2}(\{a\})$ with respect to the PDF model parameters $\{a\}$
is equivalent to minimizing $\chi'^{2} (\{a\},\{r\})$ with respect to both
$\{a\}$ and $\{r\}$. This procedure provides a much more streamlined way to
obtain the best PDF fit.  This formalism, quoted in the main text as
Eq.~(\ref{eq:SysChi1}), forms the basis of our global analysis.

Assuming the measurements {$D_i$,\ $\alpha_i$ and $\beta_{ki}$} are in accord
with normal statistics, the $\chi^{2}$ function defined by
Eq.~(\ref{eq:chi2corr}) (or (\ref{eq:Chi2sys})) should have a standard
probabilistic interpretation for a chi-squared distribution with $N_{e}$
degrees of freedom, for each experiment.

\subsection{Useful tools for evaluating fits and interpreting results}

\label{sec:Tools}

Even when the real experimental errors do not behave
in the textbook manner, as is often the case,
the formalism developed above provides useful tools for
evaluating the quality of the fits and interpreting their results. It also
provides hints on reasonable ways to deal with less-than-ideal cases. We
describe some of these tools which are used in the presentation of our fit
results throughout this paper, and in the error analyses to be discussed in
subsequent sections.

First, for any fit, Eq.~(\ref{eq:r=}) provides a best estimate of each of the
systematic errors $\{r_{k}, k=1 \dots, K\} $, which we shall denote by
$\{\hat{r}_{k}\}$. For each $k$, the parameter $\hat{r}_{k}$ should be of
order $1$, because the probability for $\hat{r}_{k}$ to be $\gg 1$ by a
random fluctuation is small. If any $\hat{r}_{k}$ turns out to be noticeably
large, the estimate of the systematic errors $\{\beta_{ki}\}$ is suspect.

Second, once the parameters that minimize the $\chi^2$ functions have been
determined, we can streamline the comparison of the fit with data---thus
evaluate the quality of the fit---by rewriting Eqs.\ (\ref{eq:Chi2sys}) and
(\ref{eq:chi2corr}) as follows:
\begin{equation}
\hat{\chi}^{2} \equiv \chi^{2}(\{\hat{a}\}) =\chi
'^{2}(\{\hat{a}\},\{\hat{r}\}) = \sum_{\mathrm{expt.}}\ \left[
\sum_{i=1}^{N_{e}}
\frac{\left(\widehat{D}_{i}-T_{i}\right)^{2}}{\alpha_{i}^{2}} +\sum_{k=1}^{K}
\hat{r}_{k}^{2}\right], \label{eq:Chi2sys1}
\end{equation}
where $\{\hat{a}\}$ (like $\{\hat{r}\}$ above), are the PDF parameters
$\{a\}$ at the $\chi^2$ minimum, and
\begin{equation}
\widehat{D}_{i} \equiv D_{i}-\sum_{k=1}^{K} \hat{r}_{k}\beta_{ki},
\label{eq:DiCorr}
\end{equation}
are the data points adjusted by the systematic errors that give rise to the
best fit.

Equation (\ref{eq:Chi2sys1}) has the simple appearance of the naive
$\chi_0^2$ function (\ref{eq:Chi2s}); but it is precise. We note: (i) The
systematic shifts of the data points associated with the fit are absorbed
into $\{\widehat{D}_{i}\}$. (ii) The denominator $\alpha_{i}^{2}$ consists of
uncorrelated errors only. (iii) The additional (last) term on the right-hand
side is just a ``constant'' when we compare the fit with data, since it is
independent of the index $i$. When the theory curves (obtained from
\{$T_{i}$\}) are compared to the adjusted data points \{$\widehat{D}_{i}$\}
to assess the goodness-of-fit, one can regard the uncorrelated errors
{$\alpha_{i}$} as the \emph{only} measurement error---clearly a great
simplification because these errors are random. Figures prepared in this way
will give a much more truthful picture of the quality of the fit than
comparing theory directly with $\{D_i\}$, since effects due to the unseen
correlated systematic errors are impossible to visualize.

\subsection{Error analysis of DIS data}

\label{app:DisErr}

The DIS experiments form the bedrock of global analysis of parton
distributions. It is seen in Table \ref{tbl:ChiAll} that the BCDMS and H1
data are within normal statistical expectations, having $\chi^{2}/N\sim 1$.
The ZEUS data have a marginally larger-than-expected $\chi^{2}/N$, of
$267/229$. For $229$ data points, the probability that $\chi^{2}\geq 267$ is
$0.063$. The NMC data has a much larger $\chi^{2}/N$ of $305/201$. This
corresponds to a nominal probability of very small magnitude indeed---$3.1
\times 10^{-6}$. (This fact was first observed in an early GKK study
\cite{GKK0}.) It is important to understand the reasons for these numbers and
to determine whether the fit to these data sets is acceptable. In addition,
in order to assess the uncertainty range of the global analysis, we will need
to adopt some uniform procedure for evaluating probabilities among the
experiments (cf.\ Sec.\ \ref{app:tolerance}). To both these ends, it is
useful to look a little deeper into the $\chi^2$ values for the individual
experiments, utilizing the tools developed in the two preceding subsections.
We use ZEUS as the first example, since its $\chi^2$ value is at the boundary
of being ``normal''.

\paragraph{ZEUS data on $\mathbf{F_{2p}(x,Q^{2})}$:}

\label{app:Zeus}

Figure \ref{fig:ZEUS} in Sec.~\ref{sec:CompData} shows the CTEQ6M
$F_{2p}(x,Q^{2})$ (solid curves) compared to the ZEUS data (corrected by the
systematic errors determined by the fit). As noted there, the agreement
between theory and data looks quite good. In order to understand the
larger-than-expected $\chi^2$ value mentioned above, we need to look deeper.
Some insight is provided by Fig.~\ref{fig:ZEUSpull}. Part (a) of this figure
shows a histogram of
\begin{equation}  \label{eq:Delta}
\Delta_{i}\equiv \frac{\widehat{D}_{i}-T_{i}}{\alpha_{i}}
\end{equation}
for the ZEUS data. The curve is a Gaussian of width $1$ with integral equal
to $N=229$, the number of data points. This histogram is the so-called ``pull
distribution''---the difference between data and theory in units of the
statistical error.%
\figZEUSpull %
The experimental fluctuations seen in Fig.~\ref{fig:ZEUSpull}a appear (i)
Gaussian, (ii) peaked near 0, and (iii) of the normal width. Considering that
there are 10 different sources of systematic errors, this test of statistical
consistency gives some confidence in the fit.

Next, we note that the net shifts of the data points due to systematic errors
are significant, but within the range expected by normal statistics.  This is
shown in Fig.~\ref{fig:ZEUSpull}b, which is similar to
Fig.~\ref{fig:ZEUSpull}a but \emph{without} correcting the data points for
systematic errors. A displacement of $\lesssim 0.5$ units in $\Delta$ is
clearly seen, but that is not unreasonably large. A detailed account of the
corrections $\{\hat{r}_{1}, \hat{r}_{2}, \dots, \hat{r}_{K}\}$ (with $K=10$
for this ZEUS data set) is shown in the following table:

\begin{center}
\begin{tabular}{|c|cccccccccc|}
\hline $k$ & 1 & 2 & 3 & 4 & 5 & 6 & 7 & 8 & 9 & 10 \\ \hline $\hat{r}_{k}$ &
$1.67$ & $-0.67$ & $-1.25$ & $-0.44$ & $-0.00$ & $-1.07$ & $1.28$ & $0.62$ &
$-0.40$ & $0.21$ \\ \hline
\end{tabular}
\end{center}

\noindent Assuming that each systematic error has a Gaussian distribution
with the published standard deviation, the probability distribution of
$r_{k}$ should be $P(r)=e^{-r^{2}/2}/\sqrt{2\pi}$. None of the corrections
listed in the table is far outside this distribution. We conclude that the
global fit is consistent with the experimental systematic errors, and the
corrections calculated from Eq.~(\ref{eq:r=}) are reasonable.

\paragraph{NMC data on
$\mathbf{F_{2}(x,Q^{2})}$ from $\mathbf{\mu{p}}$ scattering:}

\label{app:Nmc}

The comparison of NMC data on $F_{2}(x,Q^{2})$ to the CTEQ6M fit has been
shown in Fig.~\ref{fig:DisNcFt}b, Sec.~\ref{sec:CompData}. Data and theory
appear to be in general agreement---there are no systematic patterns of
deviation. However, there are clearly outlying points, showing large
point-to-point fluctuations of individual data points around the smooth
theory curves. To quantify this observation, we again examine the pull
distribution for this data set, shown in Fig.~\ref{fig:NMCpull} in the same
form as Fig.~\ref{fig:ZEUSpull} for the ZEUS data. Figure \ref{fig:NMCpull}a
shows that (i) the measurement fluctuations appear fairly Gaussian, (ii) the
distribution is peaked around $\Delta=0$, but (iii) the width of the actual
histogram appears broader than the normal distribution---there is an excess
in outlying (large fluctuation) points, and a corresponding depletion in
central (small fluctuation) points.%
\figNMCpull %

Given the larger-than-normal fluctuations, how do the systematics of the fit
measure up? Comparing Fig.~\ref{fig:NMCpull}a and Fig.~\ref{fig:NMCpull}b,
which shows the fluctuations of the uncorrected data, indicates that the net
shift in $\Delta$ due to systematic errors is $\sim 0.3$, which is quite
reasonable. The separate shifts associated with the 11 distinct systematic
errors for this fit are:

\begin{center}
\begin{tabular}{|c|ccccccccccc|}
\hline $k$ & 1 & 2 & 3 & 4 & 5 & 6 & 7 & 8 & 9 & 10 & 11 \\ \hline
$\hat{r}_{k}$ & $0.67$ & $-0.81$ & $-0.35$ & $0.25$ & $0.05$ & $0.70$ &
$-0.31$ & $1.05$ & $0.61$ & $0.26$ & $0.22$
  \\ \hline
\end{tabular}
\end{center}

\noindent
We see that the individual deviations are all reasonable, just like
the case for ZEUS.

The size of the fluctuations of the data points has very little to do with
the viability of the theory model---the excess $\chi^2$ due to such
fluctuations cannot be reduced by any improvement in the theory which
produces smooth predictions.  The quality of the fit is tied much more
closely with the systematics, which appear quite normal in the above
comparison. As observed before, in Sec.~\ref{sec:CompData}, the most
noticeable fluctuations for the NMC data---points with almost the same
$(x,Q)$ values---are from data sets taken at different incoming energies.
(They can be resolved experimentally, as in the combined data set \cite{NMC}
which we do not use.)  For these reasons, we consider the fit to the NMC data
to be acceptable.

This example vividly illustrates the usefulness of the method of error
analysis adopted here, compared to the traditional one based on the
covariance matrix. Equation (\ref{eq:Chi2sys1}) and the plots based on it
that we have shown allow a separate examination of the contributions to
$\chi^2$ from correlated and uncorrelated errors, which is useful for a more
informed assessment of the quality of the fit.  The traditional approach,
dealing only with the overall $\chi^2$, can be rather limiting when the
experimental errors do not conform to the ideal distributions.

\subsection{The tolerance criterion}

\label{app:tolerance}

In this section, we describe the procedure to calculate the range of
uncertainty for the parton distributions. In the Hessian approach we adopt,
this range is characterized by an overall tolerance parameter, $T$, that
specifies the acceptable neighborhood around the global $\chi^2$ minimum in
the parton parameter space by the condition $\Delta\chi^2 < T^2$.
Uncertainties of the PDF's and their physical predictions are all linearly
proportional to $T$. We arrive at a quantitative estimate of $T$ by examining
the range of overall $\chi^2$ along each of the eigenvector directions within
which a good fit to all data sets can be obtained, and then ``averaging'' the
ranges over the 20 eigenvector directions. The range of acceptable fits along
a given direction is estimated by combining the constraints placed on
acceptable fits by each individual experiment included in the fit, as
described below.

A key feature of this method (which makes the entire approach practical and
reliable) is the use of an orthonormal basis in the parton parameter space.
An important consequence is that the constant $\chi^2$ hypersurfaces are
simple spheroids. It comes as no surprise, then, that when we carry out the
mapping of the allowed ranges along the different eigenvector directions,
we find that they have the same order of magnitude,%
so that the averaging of the results to obtain an overall $T$ estimate makes
sense. For this reason, we do not need to show the details for all 20
eigenvector directions.

Let us consider the direction of Eigenvector $4$ as an example. Consider
points along this direction in the neighborhood of the global minimum,
labelled by $D$, the distance from the minimum. These points are candidate
fits. We first evaluate the acceptable ranges of fits with respect to the
\emph{individual} experiments, according to the known experimental
uncertainties. For experiment ($e$), the individual $\chi^2$ function,
$\chi^2_e$, is a quadratic function of $D$ with minimum at some value $D_e$.
Following Ref.\ \cite{Lagrange} (Sec.\ 4.1), we define the range of fits
acceptable with respect to experiment ($e$) by finding the
upper and lower bounds of $D$---denoted by $D_e^\pm$---using the criterion %
\begin{equation}
\int_{0}^{\chi _{_{e}}^{2}(D_{e}^{\pm })}P(\chi ^{2},N_{e})\;d\chi
^{2}\,=\,0.9 \label{eq:range}
\end{equation}
where $P(\chi^2,N_e)$ is the standard $\chi^2$-distribution for $N_e$ data
points. The ranges for the individual experiments obtained this way, each
shown as a line with the minimum of $\chi_e^2(D)$ marked by a dot, are
displayed in Fig.~\ref{fig:ToleranceA}.%
\figToleranceA %
For Eigenvector $4$ we see that the
strongest bounds on $D$ come from the data sets H1a (low $Q$ H1 data) and
BCDMSd (BCDMS deuteron target) on the negative $D$ side, and from the data
sets CCFRF3 (CCFR measurements of $F_{3}$) and ZEUS on the positive $D$ side.


A practical problem must be resolved in producing the results shown in
Fig.~\ref{fig:ToleranceA}: how does one obtain meaningful, and reasonably
uniform, estimates of the acceptable ranges for the various experiments when
the values of $\chi^2/N$ vary considerably among the data sets---and in both
directions, above and below the canonical value of $1.0$? (Cf.\ Table
\ref{tbl:ChiAll}.) In the preceding subsection, we considered some
experiments with larger-than-normal values of $\chi^2/N$. The investigations
there suggest that the deviations from normal statistical expectations can be
attributed mainly to unexplained fluctuations. In order to obtain the
acceptable ranges for the individual experiments in a uniform way, it is
reasonable to calculate the bounds $D_e^\pm$ for experiment $e$ using the
renormalized variable $\chi_{e}^{2}/[\chi_{e}^{2}(0)/N_e]$, assuming that it
obeys a chi-squared distribution with $N_{e}$ degrees of freedom. Here
$\chi_{e}^{2}(0)$ is the $\chi^2$ of experiment $e$ for the standard fit
$S_0$. This procedure provides a pragmatic and uniform way to deal with the
problems of acceptability and compatibility among the experimental input
encountered in the global analysis. It is often used in other situations
involving reconciling data from different experiments \cite{PDG}.

Eigenvector 4 was chosen for illustration purposes. The other cases are
similar. Figure \ref{fig:ToleranceB} shows results for the uncertainty ranges
of the experiments along a different direction in the parton parameter
space---that of Eigenvector 18. In this case the strongest bounds on the
displacement come from the DY and jet experiments.%
\figToleranceB %


The next problem is to convert the individual ranges in
Fig.~\ref{fig:ToleranceA} or Fig.~\ref{fig:ToleranceB} into a single
uncertainty measure. There is no unique way to do this, because the
individual ranges shown in these plots are not statistically
independent---the candidate fits along the eigenvector direction are all
obtained by global fits to the full set of experimental input. In practice,
we estimate the overall acceptable range along each eigenvector direction by
using the two most restrictive experimental constraints on either side of the
minimum. In this way, we obtain the bounds $|D|\sim 10$ for both Eigenvector
4 and Eigenvector 18, indicated by the horizontal lines in
Fig.~\ref{fig:ToleranceA} and Fig.~\ref{fig:ToleranceB}. Note that these
bounds come from different experiments, since the eigenvector directions are
sensitive to different features of the parton distributions.

The above procedure for estimating the uncertainty of $D$ was previously used
in \cite{Lagrange} to estimate the uncertainties of specific physical
predictions (such as cross sections at the Tevatron and LHC colliders), using
the Lagrange multiplier method. Each physical variable corresponds to a
definite direction in the parton parameter space. The two examples shown
above, along two eigenvector directions, are typical for all the directions.
Thus, an overall estimate for the tolerance of $T=10$ follows, as an
``average'' over all directions in the parton parameter space. This $T$
corresponds to a range of variation of the global $\chi^{2}$ of
$\Delta\chi^{2}=T^{2}=100$.

Representing the uncertainties of PDF's and their physical predictions by one
single number $T$ is clearly an oversimplification; it can only be
approximate. But given the complexity of the problem, it should be equally
clear that attempting to be more precise than this would be rather
unrealistic at this stage. In addition to the practical (hence imprecise)
measures that need to be adopted to deal with the diverse experimental data
sets with nonuniform error specifications, there are additional theoretical
uncertainties not yet included in the analysis because they are not easily
quantified. In spite of these shortcomings, however, this method of
estimating uncertainties is far more systematic and quantifiable than the
{\it ad hoc} procedures that have been used in the past. The important point
is that this method, based on the established Hessian formalism, is
fundamentally sound; its current limitations are due to compromises in
implementation necessitated by experimental realities.  The implementation
can be systematically improved as both experimental and theoretical input
improve with time.

As already noted, the estimated tolerance of $T=10$ contains experimental
uncertainties only. Uncertainties of theoretical or phenomenological origin
are not included because they are difficult to quantify. They might be
significant. For instance, we have seen throughout this paper that the
parametrization of nonperturbative PDF's has a big influence on the results.
Therefore in physical applications the criterion $T=10$ must be used with
awareness of its limitations.

\section{Study of Higher Twist Effects}

\label{app:HT}

This appendix addresses the question whether
it necessary to include higher-twist contributions
in the global analysis.
It provides the background study that leads to the conclusion
mentioned in Sec.~\ref{sec:HT}, that, with the kinematic cuts
to data included in our global analysis, higher-twist terms
are not needed.

There have been several studies of the effects of higher-twist (HT)
contributions, or power-law corrections, to $F_{2}(x,Q^{2})$
\cite{MRSThighertwist,Alekhin,AlekhinKataev,Botje}. These usually place some
emphasis on describing data in the small-$Q$ range (say, $1\;\mathrm{GeV} < Q
< 2\; \mathrm{GeV}$) where higher-twist effects are expected to be
noticeable. In the global analysis of PDF's, the emphasis is different: in
order to reliably extract the \emph{universal} parton distributions, it is
desirable to focus on the twist-2 sector of PQCD, without the complications
of any process-dependent (and model-dependent) effects, such as HT. Thus, we
adopt kinematic cuts of $Q > 2\:\mathrm{GeV}$ and $W > 3.5\:\mathrm{GeV}$ to
minimize HT effects. The question then becomes whether these kinematic cuts
are
sufficient to render HT effects numerically insignificant.\footnote{%
One would like to keep the kinematic cuts as low as is practical,
in order to include as much high-statistics data as possible
(at relatively low $Q$) in the global analysis.}
This question can only be answered phenomenologically by investigating
whether the inclusion of HT terms in the theory model is needed to achieve a
satisfactory fit, and whether the added degrees of freedom lead to a
statistically significant improvement in the fit.

To perform this study, we compare the results of our standard fits with those
including HT effects. The model adopted for the higher-twist term is similar
to that used in the literature
\cite{MRSThighertwist,Alekhin,AlekhinKataev,Botje},
\begin{equation}
F_{2}(x,Q^{2})=F^{NLO}_{2}(x,Q^{2})\left(1+\frac{H(x)}{Q^{2}}\right)
\end{equation}
where
\begin{equation}
H(x)=h_{0}+h_{1}x+h_{2}x^{2}+h_{3}x^{3}+h_{4}x^{4}.
\end{equation}
The five parameters $\{h_{0},\dots,h_{4}\}$ are determined by minimizing
$\chi^{2}$, along with the other model parameters. For simplicity, and
following the earlier higher-twist studies, we have assumed that
$\{h_{0},\dots,h_{4}\}$ are the same for all DIS processes.

The table below shows the results of our higher-twist study. The first column
is the standard fit CTEQ6M, with no higher twist, i.e., $h_{k}=0$. The second
column is the best fit with higher-twist corrections. The global $\chi^{2} $
is slightly lower with the higher-twist correction (as it must be, since
there are more parameters in this fit), but the reduction $%
\Delta\chi^{2} = -23$ (with 5 extra parameters) does not represent a
real improvement of the fit
(considering our adopted tolerance). The table also lists the values of
$\chi^{2}/N$ for individual DIS experiments.
\begin{center}
\begin{tabular}{|l|cc|}
\hline & CTEQ6M & Higher-twist fit \\
\hline $\chi^{2}$ & 1954 & 1931 \\
\hline
 &  \multicolumn{2}{c|}{$\chi^{2}_e/N_e$}   \\ \hline
BCDMS proton & 1.11 & 1.07 \\
BCDMS deuteron & 1.11 & 1.03 \\
H1 A & 0.95 & 0.94 \\
H1 B & 1.02 & 1.02 \\
ZEUS & 1.15 & 1.15 \\
NMC & 1.52 & 1.50  \\ \hline
\end{tabular}
\end{center}

 Note that the
reduction of $\chi^{2}$ from inclusion of higher twist,
comes mainly from an improved fit to the BCDMS data on $\mu{p}$ and $\mu{d}$
DIS. The HERA experiments are not affected by the higher-twist correction, as
one would expect. The NMC experiment, which, like BCDMS, has data points at
low values of $Q$, is fit only slightly better by including the higher-twist
factor $H(x)$. Figure \ref {fig:HTstudy} is a graph of the optimal function
$H(x)$. It is qualitatively similar to that found in previous studies.%
\figHTstudy %


Because the NMC data has a large $\chi^{2}$ per point compared to other DIS
experiments, we have also studied whether a higher-twist function
$H_{\mathrm{NMC}}(x)$ can be found that would significantly reduce
the value of $\chi^{2}/N$ for the NMC experiment.
The best fit for the NMC data,
keeping the PDF's unchanged but optimizing the higher-twist
correction to the NMC data, has $\chi^{2}/N=1.37$.
The associated function $H_{\mathrm{NMC}}(x)$ is somewhat different
from the best global $H(x)$---more strongly negative for $x<0.5$.
However, this ``best fit'' for NMC causes substantial increase in the
$\chi^2$'s of the other DIS experiments---the total $\chi^2$ increases (from
1954 for CTEQ6M) to 2232.
Thus, introducing HT contributions does not solve the apparent
problem of larger-than-normal $\chi^2$ for the NMC experiment in the global
fit. This is not surprising, considering our relatively large cut in the
minimum value of $Q^2 = 4 \; \mathrm{GeV}^2$.

\section{Study of Parametrizations}
\label{app:Param}

We have noted in Secs.~\ref{sec:Parametrization}, \ref{sec:PdfRes} and
\ref{sec:Mrst01} that parametrization of the nonperturbative QCD PDF's now
has an important bearing on the results of the global analysis, given the
much improved experimental constraints and the newly developed theoretical
methods. We have rather extensively explored the influence of
parametrization, using the functional forms described in
Sec.~\ref{sec:Parametrization} and the iterative Hessian eigenvector method
for matching the degrees of freedom in the parametrization with applicable
experimental constraints (cf.\ Sec.~\ref{sec:PdfUnc}).

Even with tools like these, the choice of parametrization still ultimately
involves subjective (i.e., physical) judgements. In this appendix, we shall
give only one example of the studies we have carried out, relating to the
behavior of the gluon distribution at both small and large $x$, as discussed
in Sec.~\ref{sec:DelPdf} and Sec.~\ref{sec:Mrst01}.

This study was motivated by an attempt to understand the results of CTEQ6
fits described in Sec.~\ref{sec:PdfRes} in light of two issues raised by
\cite{Mrst01}: (i) Do the recent HERA data imply a negative gluon
distribution at small $x$ at the scale $Q=1\,\mathrm{GeV}$? (ii) Do good fits
to the new D\O\ jet data necessitate an artificial-looking humped structure
of the gluon at large $x$ at low $Q$? The seeming contradiction between the
observations of \cite{Mrst01}, which give rise to these questions, and the
apparent good all-around CTEQ6 fits is resolved by this exercise.


First, the observations of \cite{Mrst01} were confirmed when we performed a
global fit using the MRST2001 functional form for the gluon, but keeping our
parametrization of the quark degrees of freedom, and fitting to our full set
of data with our definition (\ref{eq:chi2corr}) for the global $\chi^2$.
Figure \ref{fig:MrstFit}a shows the resulting gluon distribution at $Q=1$,
$2$, $5$, $100 \, \mathrm{GeV}$. At the scale $Q=1 \; \mathrm{GeV}$, both the
negative gluon at small $x$ and the humped structure at large $x$ found by
the least-$\chi^2$ MRST2001J fits are reproduced. For comparison, Figure
\ref{fig:MrstFit}b shows the same distributions from a fit using our
parametrization.  (To make the comparison possible, this fit was done with
$Q_0 = 1 \, \mathrm{GeV}$ in place of the CTEQ6 value $Q_0 = 1.3 \,
\mathrm{GeV}$.) In this case, there is only a slight shoulder at large $x$
for $Q = 1 \, \mathrm{GeV}$, and the distribution has become completely
smooth by $Q = 2 \, \mathrm{GeV}$.  Hence the hump structure seen by MRST is
an artifact of the particular choice of the parametrization. The gluon
distribution is constrained to be positive definite in this parametrization.
These two fits differ in overall $\chi^2$ by an amount that is less than our
tolerance estimate, so we do not find convincing evidence for a negative
gluon distribution even at the very low scale of $Q=1 \; \mathrm{GeV}$, where
the parton distributions are ambiguous anyway because of higher-order and
higher-twist corrections.

\newpage%
\figMrstFit %


Since there is no firm theoretical requirement for a positive-definite gluon
distribution at $Q$ as low as $1 \, \mathrm{GeV}$, we also tried an
alternative parametrization for the gluon in which the CTEQ6 functional form
is multiplied by an additional factor $(1 + A_6/x)$ that allows $G(x,Q_{0})$
to go negative at small $x$. The resulting gluon distributions are shown in
Fig.~\ref{fig:NegFit}a.%
\figNegFit %
This fit has one more fitting parameter, and hence it
results in a slightly lower overall $\chi^2$ than CTEQ6M, but the reduction
is well within our tolerance range. An interesting feature of this fit is, of
course, that it does become negative at $Q=1 \, \mathrm{GeV}$. But, like the
case in Fig.~\ref{fig:MrstFit}a, the distribution rapidly becomes positive
under QCD evolution. No trace of the negative region is seen at $Q=2\;
\mathrm{GeV}$. Is the seeming preference for a negative gluon at $Q=1\;
\mathrm{GeV}$ physically significant? We can answer this question
quantitatively by mapping out the range of uncertainty of the gluon
distribution at $Q=1\, \mathrm{GeV}$ using the Hessian method. The result is
shown in Fig.~\ref{fig:NegFit}b. The range of uncertainty, given current
experimental constraints, is very large at the $Q=1\; \mathrm{GeV}$ scale,
and it covers both positive and negative territories at small $x$, as well as
large $x$! Due to the nature of QCD evolution, the uncertainty decreases
rapidly with increasing $Q$, as shown in Fig.\ \ref{fig:SmlGluUn2}.%
\figSmlGluUnTwo %


\end{document}